\documentclass[twocolumn]{fairmeta}

\usepackage{amsmath,amsthm,amssymb,bm, mathtools}
\usepackage{amsfonts}
\usepackage{thmtools} 
\usepackage{bbm}
\usepackage{algorithm}
\usepackage{algorithmic}

\usepackage{float}
\newfloat{algorithm}{t}{lop}
\usepackage{enumitem}

\usepackage{pifont}%

\newcommand{\cev}[1]{\reflectbox{\ensuremath{\vec{\reflectbox{\ensuremath{#1}}}}}}

\newenvironment{talign*}
 {\csname align*\endcsname}
 {\endalign}

\usepackage{xspace}

\definecolor{cadetblue}{rgb}{0.37, 0.62, 0.63}
\definecolor{mygray}{gray}{0.95}

\makeatletter
\DeclareRobustCommand{\cev}[1]{%
  {\mathpalette\do@cev{#1}}%
}
\newcommand{\do@cev}[2]{%
  \vbox{\offinterlineskip
    \sbox\z@{$\m@th#1 x$}%
    \ialign{##\cr
      \hidewidth\reflectbox{$\m@th#1\vec{}\mkern4mu$}\hidewidth\cr
      \noalign{\kern-\ht\z@}
      $\m@th#1#2$\cr
    }%
  }%
}
\makeatother

\usepackage{microtype}
\usepackage{graphicx}
\usepackage{booktabs} %
\usepackage{adjustbox}
\usepackage{xcolor}
\usepackage{multirow}

\usepackage{hyperref}

\definecolor{mydarkblue}{rgb}{0,0.3,0.6}
\definecolor{es-blue}{rgb}{0.165, 0.616, 0.957}
\newcommand{\ourmodel}{eSEN}

\usepackage[noabbrev,nameinlink,capitalize]{cleveref}
\theoremstyle{plain}  %

\theoremstyle{definition}  %

\title{Learning Smooth and Expressive Interatomic Potentials for Physical Property Prediction}

\author[1]{Xiang Fu}
\author[1]{Brandon M.\ Wood}
\author[1]{Luis Barroso-Luque}
\author[1]{Daniel S.\ Levine}
\author[1]{Meng Gao}
\author[1]{Misko Dzamba}
\author[1]{C.~Lawrence Zitnick}

\affiliation[1]{Fundamental AI Research (FAIR) at Meta}

\abstract{
Machine learning interatomic potentials (MLIPs) have become increasingly effective at approximating quantum mechanical calculations at a fraction of the computational cost. However, lower errors on held out test sets do not always translate to improved results on downstream physical property prediction tasks. We propose testing MLIPs on their practical ability to conserve energy during molecular dynamic simulations. If passed, improved correlations are found between test errors and their performance on physical property prediction tasks. We identify choices which may lead to models failing this test, and use these observations to improve upon highly-expressive models. The resulting model, \ourmodel, provides state-of-the-art results on a range of physical property prediction tasks, including materials stability prediction, thermal conductivity prediction, and phonon calculations.
}

\correspondence{Xiang Fu (\email{xiangfu@meta.com}) and C.~Lawrence Zitnick (\email{zitnick@meta.com})}

\metadata[Code]{\url{https://github.com/facebookresearch/fairchem}}
\metadata[Checkpoints]{\url{https://huggingface.co/facebook/OMAT24}}

\begin{document}

\maketitle

\section{Introduction}
\label{sec:intro}

\begin{figure}[t]
\includegraphics[width=\linewidth]{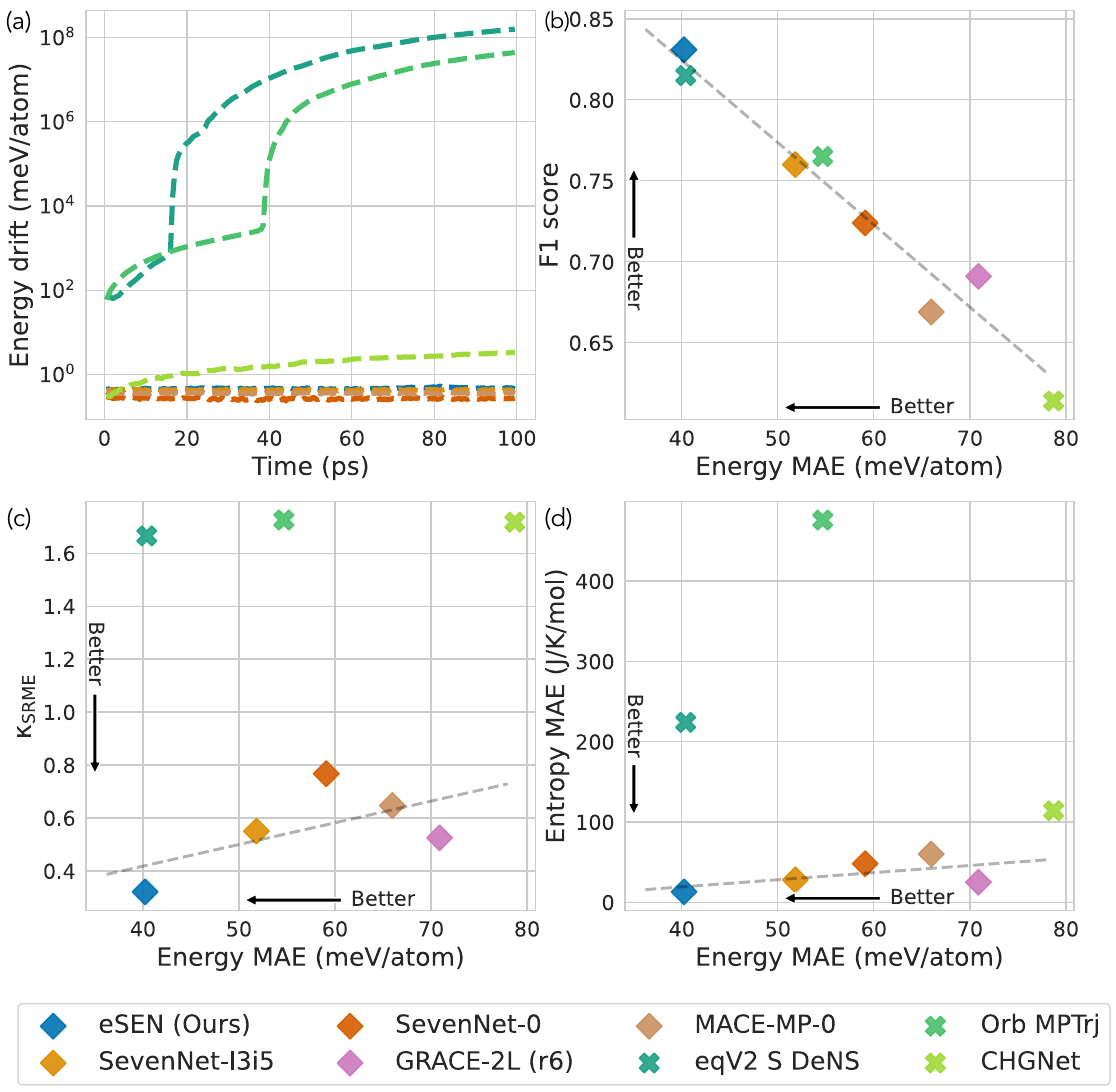}
\caption{(a) Energy conservation in MD simulations. Direct-force models (Orb, eqV2) and CHGNet fail to conserve. (b) A higher F1 score on the Matbench-Discovery strongly correlates with a lower test-set energy MAE. (c) Test-set energy MAE and $\kappa_{\mathrm{SRME}}$ on the Matbench-Discovery benchmark. (d) Test-set energy MAE and vibrational entropy MAE on the MDR Phonon benchmark. Our model (\ourmodel) achieves the best performance on all benchmarks. A higher correlation between test-set energy MAE and physical property prediction performance can be observed among energy-conserving models. All models are trained on MPTrj.}
\label{fig:overview}
\end{figure}

Density Functional Theory (DFT), which models the electrons in materials and molecules, serves as the foundation for many modern drug and materials discovery workflows. Unfortunately, DFT calculations are notoriously computationally intensive, scaling cubically with the number of electrons in the system:  $O(n^3)$. Machine learning interatomic potentials (MLIPs) are promising in approximating and expediting DFT calculations. With increasing data set sizes and model innovations, MLIPs have shown substantial improvements in accuracy and generalization capabilities~\citep{batatia2023foundation, merchant2023scaling, yang2024mattersim, barroso2024open}. 

Predicting physical properties in chemistry and materials science often requires complex workflows involving numerous evaluations of DFT or MLIPs. For example, in molecular dynamics (MD) simulations, forces are predicted over thousands to millions of time steps. However, the MLIP literature has mostly focused on assessing models based on energy and force predictions over static DFT test sets rather than directly assessing their performance in complex simulations. This approach has limitations, as improved accuracy on test sets does not always lead to better predictions of physical properties~\citep{pota2024thermal, loew2024universal}. 

In this paper, we address two questions: Why does higher test accuracy sometimes fail to enhance a model's ability to predict physical properties, and how can we improve MLIPs to excel in this area? We first outline four critical property prediction tasks and identify the properties required for an MLIP to succeed in these tasks. These properties entail learning a conservative model with continuous and bounded energy derivatives, indicating a smoothly-varying and physically meaningful energy landscape. To test whether these properties hold, we propose testing the ability of MLIPs to practically conserve energy in MD simulations. We demonstrate models that pass this test have a higher correlation between test errors and property prediction accuracy.  

Building on these insights, we present a novel MLIP called \ourmodel~and training approach that achieves state-of-the-art (SOTA) performance on complex property prediction tasks. Specifically, our model is capable of running energy-conserving MD simulations for out-of-distribution systems (\cref{fig:overview}~(a)). For materials stability prediction, \ourmodel~achieves a leading F1 score of $0.831$ and a $\kappa_{\mathrm{SRME}}$ of $0.340$ on the compliant Matbench-Discovery benchmark~\citep{riebesell2023matbench, pota2024thermal}. Previous models are only able to excel in one of these metrics (~\cref{fig:overview}~(b,c)). We also achieve a SOTA F1 score of $0.925$ and $\kappa_{\mathrm{SRME}}$ of $0.170$ on the non-compliant category. On the MDR Phonon benchmark~\citep{loew2024universal}, SOTA results are found (\cref{fig:overview}~(d)). Finally, \ourmodel~achieves the highest test accuracy on the SPICE-MACE-OFF dataset~\citep{kovacs2023mace}.

\section{Preliminaries}
\label{sec:related}

\subsection{Machine learning interatomic potentials} 

Under the Born-Oppenheimer approximation~\citep{oppenheimer1927quantentheorie} utilized by DFT~\citep{parr1979local}, the Potential Energy Surface (PES) can be written as a function of positions, $\bm r$, and atomic numbers, $\bm a$: $E(\bm r, \bm a)$. Per-atom forces can be calculated by taking the negative gradient of the PES with respect to the atom positions, $\bm F = -\nabla_{\bm r} E$. For periodic systems such as inorganic materials, the lattice parameters $\bm l$ are also considered ($E(\bm r, \bm a, \bm l)$), and the stress $\bm \sigma$ may also be calculated, which can be understood as the gradient of the potential energy surface with respect to the lattice parameters. 

The goal of an MLIP~\citep{unke2021machine} is to predict the exact same properties as DFT from a training dataset of DFT calculations~\citep{oc20, riebesell2023matbench, loew2024universal}. The most straightforward benchmark for MLIPs is to evaluate the model on a held-out test set of DFT calculations, and compare models based on the mean absolute error (MAE) or root mean squared error (RMSE) of energies, forces, or stresses. To bridge the gap between these performance metrics and practical applicability, we need to ensure they correlate with physical property prediction tasks, such as those described next.

\subsection{Physical property prediction tasks}

\textbf{Geometry optimization/relaxation.} 
Many computational chemistry and materials science tasks rely on atomic systems being in stable configurations, which correspond to minima of the PES. Stable states are found by minimizing the potential energy using an optimization procedure that iteratively updates atom positions based on the predicted forces ($\bm F = -\nabla_{\bm r} E$). Given that many physical properties are evaluated at or near equilibrium states, geometry optimization (also referred to as ``relaxation'') is usually the first step in most computational workflows. 

\textbf{MD simulations.} 
Simulating the time evolution of atomic systems enables us to gain understanding of various chemical and biological processes, as well as enabling the calculation of macroscopic properties, such as liquid densities, that can be experimentally verified. For the task of molecular dynamics simulation, we typically use a potential to compute the per-atom forces which are then used to numerically integrate Newton’s equations of motion. In this work, we will focus on the \textbf{microcanonical ensemble (NVE)}, where the number of particles (N), the volume of the system (V), and the energy of the system (E) are kept constant. 

\textbf{Phonon and thermal conductivity calculations.}
Precise predictions of phonon band structures and vibrational modes are essential for understanding various material properties, including dynamical stability, thermal stability~\cite{bartel2022review, fultz2010vibrational}, thermal conductivity~\cite{razeghi2002thermal}, and optoelectronic behavior~\cite{ganose2021efficient}. The calculation of phonon band structures requires the MLIP to accurately predict higher-order derivatives and capture the subtle curvature of the true PES around critical points. Recent work~\cite{pota2024thermal} has demonstrated the usage of MLIPs in predicting thermal conductivity ($\kappa$) by solving the Wigner transport equation~\citep{simoncelli2022wigner}. In order to accurately predict $\kappa$, MLIPs must reliably capture both harmonic and anharmonic phonon behavior, which necessitates the calculation of second and third derivatives of the learned PES.

\section{Desideratum for physical property prediction}
\label{sec:conservation}

We begin the section by defining what it means for an MLIP to be energy conserving, which is a fundamental principle for applications such as MD simulations~\citep{tuckerman2023statistical}. For many physical property prediction tasks that probe the higher-order derivatives of the PES it is also important that the PES's derivatives are well-behaved (they exist and are bounded). To indicate whether a PES meets these criteria, we discuss how an MLIP's ability to conserve energy given fixed simulation settings may be used. 

\subsection{Conservative forces}
For a force model to be conservative, the work done by moving in a closed path must be zero, i.e., the integration of the forces along any path that starts and ends at the same point is zero:
\begin{align}
\oint \bm F \cdot d\bm r = 0
\end{align}
This property holds if the forces are calculated as the negative derivative of the PES with respect to the atom positions~\citep{unke2021machine}. However, predicting forces as derivatives requires an additional backpropagation step through the network, which increases the computational cost of the MLIP. Alternatively, some networks~\citep{liao2023equiformerv2, neumann2024orb} directly predict forces using a separate force head to increase efficiency\footnote{Strictly speaking, direct-force models are not truly ``potentials'', but rather (non-conservative) ``force fields''.}. Although direct-force models can achieve high accuracy, their non-conservative nature leads to significantly larger errors in certain property prediction tasks~\citep{fu2023forces, loew2024universal, pota2024thermal, bigi2024dark}.

\subsection{Bounded energy derivatives}

Conservative forces is a necessary but not sufficient condition for an MLIP to demonstrate energy conservation in MD. In practice, MD simulations use a finite-order numerical integration algorithm and a finite time step $\Delta t$, which introduces truncation errors. The most commonly used integrator for the NVE ensemble is the Verlet algorithm--a second-order integrator. The Verlet integrator is known to approximately conserve the total energy of the system in long-time simulations. As shown by Theorem 5.1 of \citealt{hairer2003geometric}, the total energy drift of a simulation satisfies
\begin{align}
 | E(\bm r_T, \bm a) - E(\bm r_0, \bm a) | \leq C\Delta t^2 + C_N \Delta t^N T,
 \label{eqn:bounds}
\end{align}
where $T$, $0 \leq T \leq \Delta t^{-N}$, is the total simulation time, $N$ is a positive integer representing the highest order for which the $N$th-order derivative of $E$ is continuously differentiable with a bounded derivative, and $\bm r_0$ and $\bm r_T$ are the starting and ending positions of the atoms in the simulation respectively. The constants $C$ and $C_N$ are independent of $T$ and $\Delta t$. The energy drift bound contains two terms: the first term represents a time-independent fluctuation of $O(\Delta t^2)$, and the second term represents the long-term energy conservation. The proof for this theorem is long and technical, for which we refer interested readers to \citealt{hairer2003geometric} and \citealt{hairer2006geometric} for more details.

In \cref{eqn:bounds}, the $\Delta t^N$ in the second term and the bound on the simulation time $T \leq \Delta t^{-N}$ implies that the PES must be continuously differentiable to high order for energy conservation in long-time simulations. The critical constant $C_N$ depends on the bounds of the derivatives of $E$ up to the $(N + 1)$th order. This implies that, given a fixed time step size, $E$ and its higher-order derivatives up to the $(N)$th order all need to be continuously differentiable with bounded derivatives to maintain long-time conservation. If the derivatives of a PES are more tightly bound, approximate energy conservation will be maintained even at larger step sizes $\Delta t$. Therefore, the magnitude of $\Delta t$ for which the energy is stable can be viewed as a proxy for the derivative bounds of the estimated PES. Alternatively, if a certain time step is known to be stable when using DFT, we can determine whether an MLIP has similar bounds on higher-order derivatives by testing whether it is also stable using the same time step. 

\section{\ourmodel}
\label{sec:model}

\begin{figure}[t]
\includegraphics[width=\linewidth]{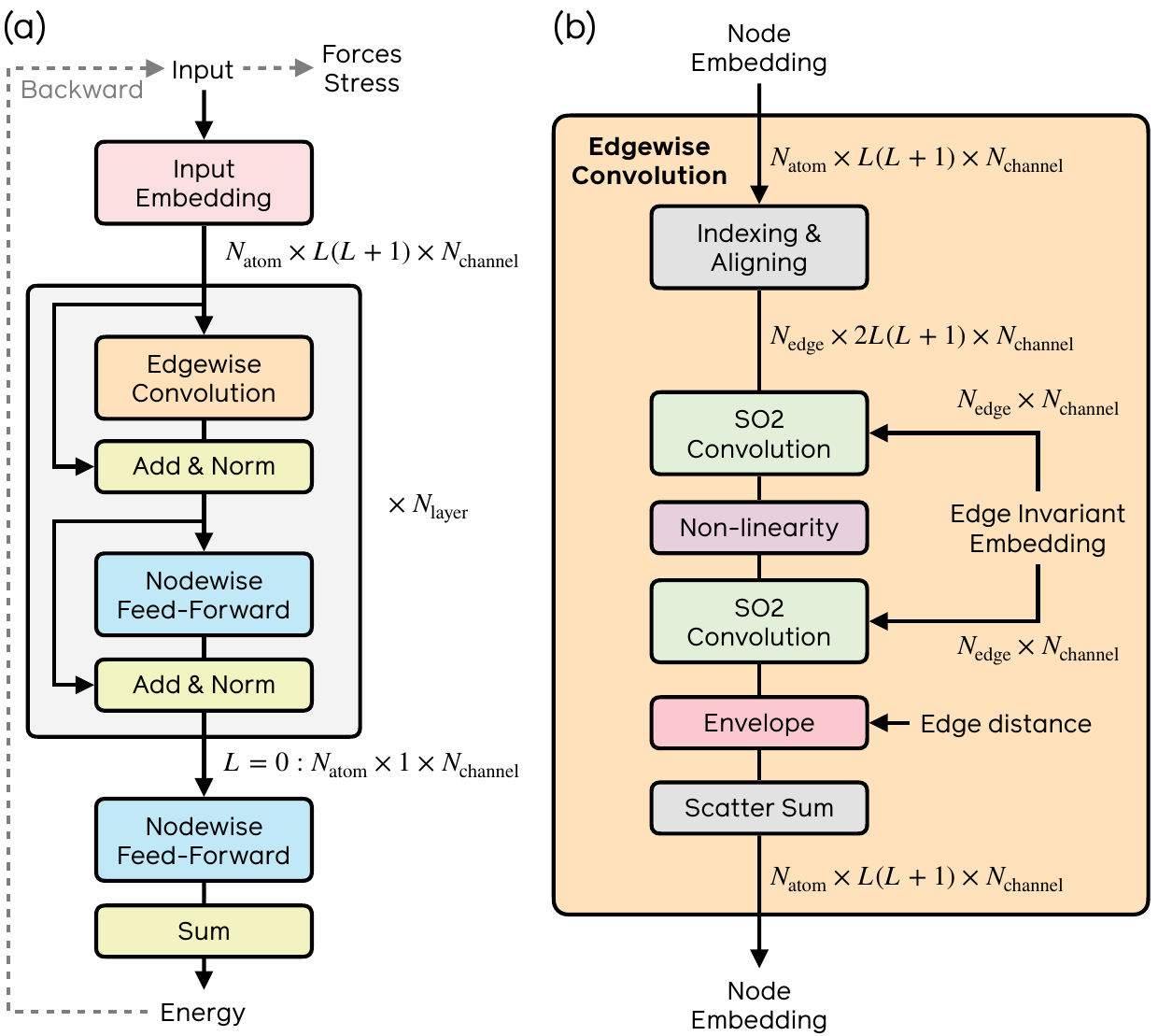}
\caption{(a) The \ourmodel\ architecture. The high-level architecture is similar to Transformer/Equiformer, while the edgewise/nodewise layers are simplified/enhanced. The final-layer $L=0$ features are used to predict nodewise energy, which is summed to get the total potential energy $E$. Forces and stress are obtained through back-propagration. (b) The \textbf{Edgewise Convolution} layer in \ourmodel.
}
\label{fig:model}
\end{figure}

We propose \textbf{\underline{e}}quivariant \textbf{\underline{S}}mooth \textbf{\underline{E}}nergy \textbf{\underline{N}}etwork (\textbf{\ourmodel}), a new MLIP architecture that improves upon architectures that demonstrate high test accuracies to achieve effective physical property predictions. \ourmodel\ is a message-passing neural network that conducts multiple blocks of edgewise and nodewise neural processing. Initially, all nodes are embedded as multi-channel spherical harmonic representations. Each \ourmodel\ layer block updates the node embedding by conducting an edgewise convolution, followed by a nodewise feed-forward network with normalization layers and residual connections between all layers. 

A model diagram is shown in \cref{fig:model}. \ourmodel\ utilizes the same SO2 convolution layer from the equivariant spherical channel network (eSCN) architecture \cite{passaro2023reducing} inside the edgewise convolution block. Compared to eSCN, our edgewise convolution blocks first concatenate the source and target node embedding, then apply two SO2 convolution layers with an intermediate non-linearity. We also add an envelope function (details in \cref{sec:design}) which is not in eSCN. The nodewise feed-forward layer uses two equivariant linear layers and an intermediate SiLU-based gated non-linearity~\citep{weiler20183d, geiger2022e3nn}, which is the same as Equiformer~\citep{liao2022equiformer}. Unlike eSCN and EquiformerV2~\citep{liao2023equiformerv2}, which projects the spherical-harmonics channels onto spatial grids for nodewise processing, the nodewise layers in \ourmodel\ do not discretize the node representations. As we demonstrate in \cref{sec:design}, this design improves the ability of the model to conserve energy. Normalization is performed using the equivariant layer normalization~\citep{ba2016layer} proposed by Equiformer~\citep{liao2022equiformer}. In the next section, we conduct an in-depth analysis of the key design choices for energy conservation, which we argue is important for accurate physical property prediction.

\section{Design choices for enhancing physical property prediction}
\label{sec:design}

As discussed in \cref{sec:conservation}, having conservative forces with continuous and bounded energy derivatives are properties an MLIP should obey for MD simulations. It can also be seen as a prerequisite for the MLIP to accurately capture higher-order behavior of the PES and thus high accuracy in physical property prediction tasks such as phonon calculations. Motivated by this observation, we identify design choices that impact a model's ability to conserve energy and whether its PES varies smoothly. These design choices can be categorized into three aspects: (1) conservative vs. direct-force prediction; (2) discretization of the representation; and (3) obtaining a continuous and smoothly varying PES. For many of these design aspects, their impact on the desired properties is not well understood. 

To quantify whether an MLIP's PES is continuous and smoothly varying, we measure the ability of the resulting MLIP to conserve energy during MD simulations with a predetermined fixed time step. We trained \ourmodel\ models under the same hyperparameters while ablating one design choice at a time. We construct out-of-distribution (OOD) MD simulation tasks for both inorganic materials and organic molecules using models trained on the MPTrj~\citep{jain2013materials, deng2023chgnet} and the SPICE-MACE-OFF~\citep{eastman2023spice, kovacs2023mace} datasets. For inorganic materials, we compute an average conservation error over 81 NVE MD simulations of 100 ps based on the TM23 dataset's simulation settings~\citep{owen2024complexity}. For organic molecules, we compute an average conservation error over 7 NVE MD simulations of 100 ps based on the MD22 dataset's simulation settings~\citep{chmiela2023accurate}.  All \ourmodel\ models are 2-layer with 3.2M trainable parameters. We include details regarding the task protocol in \cref{appendix:experimental}. 

\subsection{Direct-force prediction}

\begin{figure*}[t]
\includegraphics[width=\textwidth]{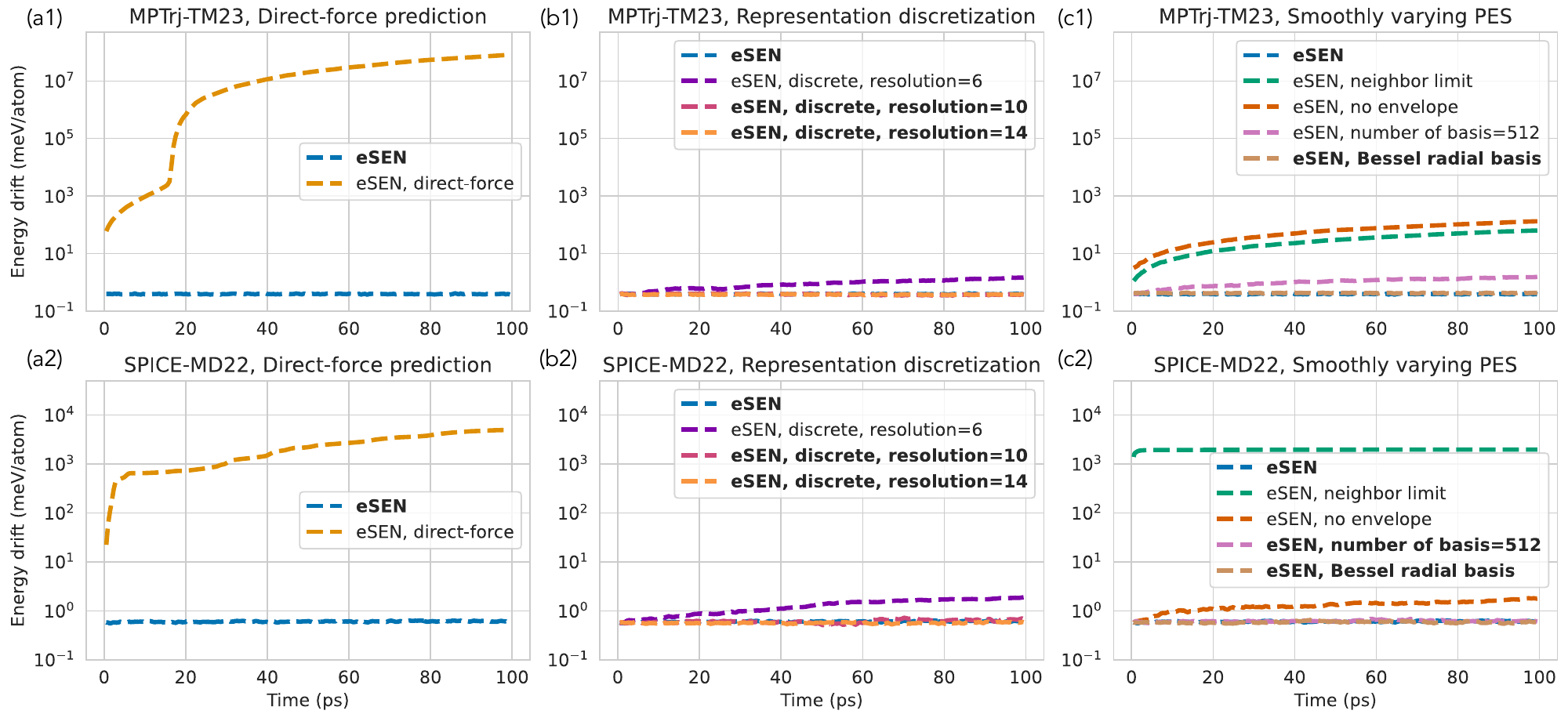}
\caption{
Conservation error on the TM23 task (top row) and MD22 task (bottom row) for ablating design choices of \ourmodel. Models that conserve energy are \textbf{bolded} in the legends.
}
\label{fig:conservation}
\end{figure*}

Models that directly predict forces $\hat{\bm F}$ from the atomic configuration may produce forces that are inconsistent with the energy prediction, i.e., $\hat{\bm F} \neq -\nabla_{\bm r} \hat{E}$, and more importantly are unlikely to be conservative. From the perspective of minimizing the test error, the direct-force approach has strong motivations: it avoids the backward pass for force prediction, which significantly improves model efficiency and enables low-precision training which further accelerates training. Empirically, current SOTA accuracy on the OC20, OC22, and Matbench-Discovery~\citep{oc20, oc22, riebesell2023matbench} benchmarks are achieved by direct-force models. Despite this, the direct-force formulation results in significant energy drift in MD simulations, as shown in \cref{fig:conservation}~(a1, a2). For this reason, we compute forces as the negative gradient
of the PES with respect to the atom positions in \ourmodel.

\begin{figure}[h!]
\includegraphics[width=\linewidth]{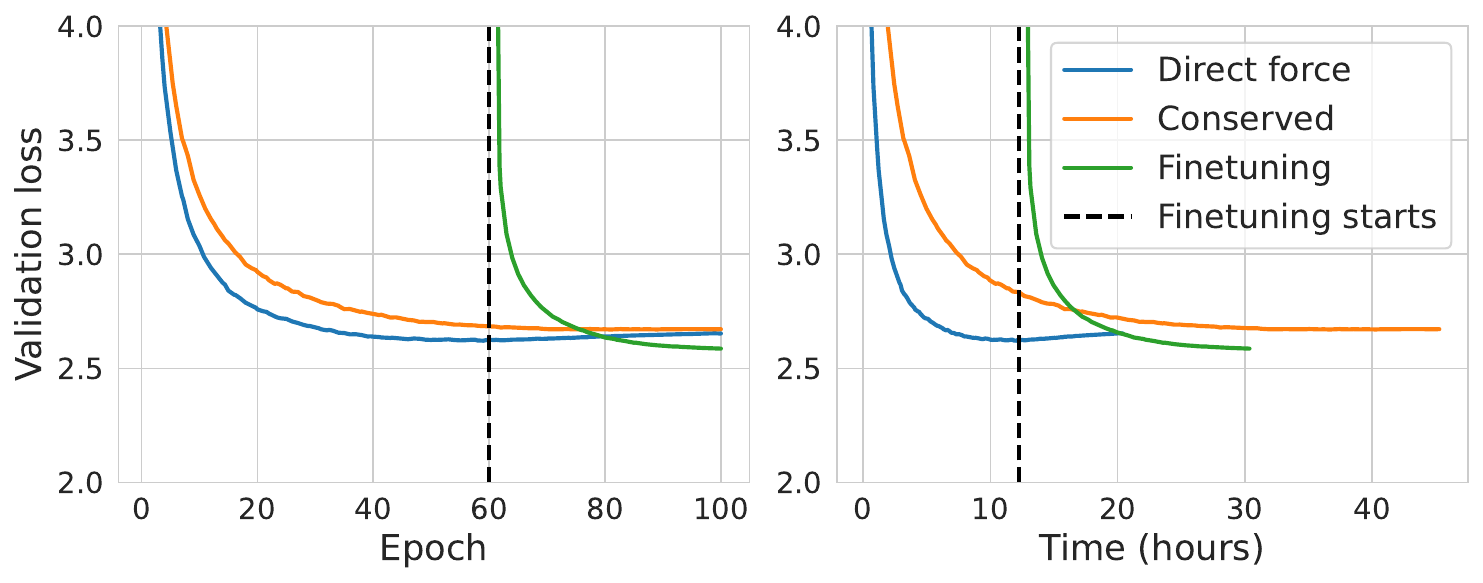}
\caption{
Validation loss curves for epoch and wallclock time. 
}
\label{fig:finetuning}
\end{figure}

\textbf{Direct-force pre-training.} Although direct-force models are not suitable for certain physical property prediction tasks, they may still offer advantages~\citep{bigi2024dark,amin2025towards}. We demonstrate their efficiency can offer significant benefit as a pre-training strategy for a conservative model. \cref{fig:finetuning} shows the validation loss of 2-layer \ourmodel\ models trained on the MPTrj dataset: direct-force, conservative, and conservative fine-tuning from a pre-trained direct-force backbone. We start from a direct-force model trained for 60 epochs, remove its direct-force prediction head, and fine-tune using conservative force prediction. The conservative fine-tuned model achieves a lower validation loss after being trained for 40 epochs compared to the from-scratch conservative model being trained for 100 epochs. The fine-tuning strategy also reduces the wallclock time for model training by 40\%. The strategy of combining direct-force pre-training and conservative fine-tuning is also shown to be effective under different data/model settings~\citep{bigi2024dark}.

\subsection{Representation discretization}
As proposed by \citealt{cohen2016group} and later used in eSCN and EquiformerV2~\citep{zitnick2022spherical, passaro2023reducing, liao2023equiformerv2}, non-linearities may be performed by projecting the spherical harmonics to a discrete grid. A $1 \times 1$ convolution or pointwise non-linearity may then be applied to this grid, which then get projected back to the spherical-harmonics space. The non-linear step may introduce higher-frequency signals than cannot be properly represented by the spherical harmonics, i.e., they are beyond the Nyquist frequency. This can lead to sampling errors that break strict equivariance and energy conservation. This problem can be mitigated by sampling the grid at higher resolutions as shown in \cref{fig:conservation}~(b1, b2). In \ourmodel, we instead use the SiLU-based equivariant Gated non-linearity~\citep{weiler20183d, geiger2022e3nn} that performs the non-linearity directly in the spherical harmonic representation. This does not require a projection to a discrete grid, so the model is perfectly equivariant and conservative up to numerical accuracy. 

\subsection{Smoothly varying PES}

Subtle choices in the design of MLIPs can have a significant impact on whether a PES varies smoothly and can even lead to the presence of discontinuities. These include how neighboring atoms are chosen, whether envelope functions are used near atom distance cutoffs, and which basis functions are used to embed pairwise atom distances. We discuss each of these in turn. 

\textbf{A maximum number of neighbors limit} in graph construction has been found to improve training efficiency without compromising test error~\citep{liao2023equiformerv2, qu2024the}. However, it results in a discontinuity in the learned PES as the nearest-K neighbors may change drastically under a small perturbation of the atom positions. As shown in \cref{fig:conservation}~(c1, c2), having a maximum neighbor limit breaks energy conservation. In \ourmodel, instead of limiting the number of neighbors, we use the common approach of applying a distance cutoff ($6\textup{\AA}$) under which all neighbors are kept.

\textbf{Envelope functions} were first introduced in the DimeNet architecture~\citep{gasteiger2020directional} to improve model smoothness. The radial basis function used in MLIPs is not twice continuously differentiable due to the use of a finite cutoff during graph construction. By applying a polynomial envelope function on the edge messages, the values in an edge message and its first/higher-order derivatives with respect to atom positions decays to 0 when the edge distance approaches the cutoff distance. \cref{fig:conservation}~(c1, c2) shows a model fails to conserve energy without the envelope function.

\begin{table}[t]\centering
\caption{
Test set MAE for design choices studied in \cref{sec:conservation}. The conserved model significantly outperforms the direct-force model on SPICE-MACE-OFF. $N_{\mathrm{basis}}=512$ performs slightly better on SPICE but slightly worse on MPTrj. Other \ourmodel\ variants all have similar test errors on MPTrj/SPICE-MACE-OFF. Energy MAE is in meV/atom. Force MAE is in meV/\AA. Stress MAE is in meV/\AA/atom.
\label{tab:design_test_error}
}
\begin{adjustbox}{width=\linewidth}
\begin{tabular}{l|ccc|cc}
\toprule
 & \multicolumn{3}{c|}{\textbf{MPTrj}} & \multicolumn{2}{c}{\textbf{SPICE}} \\
\textbf{Model} & \textbf{Energy} & \textbf{Force} & \textbf{Stress} & \textbf{Energy} & \textbf{Force} \\
\midrule
\ourmodel & 17.02 & 43.96 & 0.14 & 0.23 & 6.36 \\
\ourmodel, direct & 18.66 & 43.62 & 0.16 & 0.56 & 10.98 \\
\ourmodel, neighbor limit & 17.30 & 44.11 & 0.14 & 0.24 & 6.52 \\
\ourmodel, no envelope & 17.60 & 44.69 & 0.14 & 0.23 & 6.33 \\
\ourmodel, $\mathrm{N}_{\mathrm{basis}}=512$ & 19.87 & 48.29 & 0.15 & 0.19 & 5.40 \\
\ourmodel, Bessel & 17.65 & 44.83 & 0.15 & 0.20 & 5.54 \\
\ourmodel, discrete, res=6 & 17.05 & 43.10 & 0.14 & 0.26 & 6.34 \\
\ourmodel, discrete, res=10 & 17.11 & 43.13 & 0.14 & 0.33 & 6.57 \\
\ourmodel, discrete, res=14 & 17.12 & 43.09 & 0.14 & 0.33 & 6.51 \\
\bottomrule
\end{tabular}
\end{adjustbox}
\end{table}

\begin{table*}[!htp]\centering 
\caption{
Matbench-Discovery benchmark results of compliant models (trained only on MPtrj or its subset) with results on the unique prototype split. MAE is in units of eV/atom. ($\uparrow$/$\downarrow$) stands for higher/lower the better.
\label{tab:mbd}
}
\begin{adjustbox}{width=\textwidth}
\begin{tabular}{l|ccccccccccc}\toprule
Metric & eSEN-30M-MP & eqV2 S DeNS & MatRIS-MP & AlphaNet-MP & DPA3-v2-MP & ORB v2 MPtrj & SevenNet-l3i5 & GRACE-2L-MPtrj & MACE-MP-0 & CHGNet & M3GNet \\\midrule
F1 $\uparrow$ & \textbf{0.831} & 0.815 & 0.809 & 0.799 & 0.786 & 0.765 & 0.760 & 0.691 & 0.669 & 0.613 & 0.569 \\
DAF $\uparrow$ & \textbf{5.260} & 5.042 & 5.049 & 4.863 & 4.822 & 4.702 & 4.629 & 4.163 & 3.777 & 3.361 & 2.882 \\
Precision $\uparrow$ & \textbf{0.804} & 0.771 & 0.772 & 0.743 & 0.737 & 0.719 & 0.708 & 0.636 & 0.577 & 0.514 & 0.441 \\
Accuracy $\uparrow$ & \textbf{0.946} & 0.941 & 0.938 & 0.933 & 0.929 & 0.922 & 0.920 & 0.896 & 0.878 & 0.851 & 0.813 \\
MAE $\downarrow$ & \textbf{0.033} & 0.036 & 0.037 & 0.041 & 0.039 & 0.045 & 0.044 & 0.052 & 0.057 & 0.063 & 0.075 \\
R2 $\uparrow$ & \textbf{0.822} & 0.788 & 0.803 & 0.745 & 0.804 & 0.756 & 0.776 & 0.741 & 0.697 & 0.689 & 0.585 \\
\midrule
$\kappa_{\mathrm{SRME}}$ $\downarrow$ & \textbf{0.340} & 1.676 & 0.861 & 1.31 & 0.959 & 1.725 & 0.550 & 0.525 & 0.647 & 1.717 & 1.412 \\
RMSD $\downarrow$ & \textbf{0.0752} & 0.0757 & 0.0773 & 0.1067 & 0.0823 & 0.1007 & 0.0847 & 0.0897 & 0.0915 & 0.0949 & 0.1117 \\
\bottomrule
\end{tabular}
\end{adjustbox}
\end{table*}

\begin{table*}[!htp]\centering 
\caption{
Matbench-Discovery benchmark results of non-compliant models with results on the unique prototype split.
\label{tab:mbd_nc}
}
\begin{adjustbox}{width=\textwidth}
\begin{tabular}{l|ccccccccc}\toprule
Model & eSEN-30M-OAM & eqV2-M-OAM & ORB v3 & SevenNet-MF-ompa & DPA3-v2-OpenLAM & GRACE-2L-OAM & MatterSim-v1-5M & MACE-MPA-0 & GNoME \\\midrule
F1 $\uparrow$ & \textbf{0.925} & 0.917 & 0.905 & 0.901 & 0.890 & 0.880 & 0.862 & 0.852 & 0.829 \\
DAF $\uparrow$ & \textbf{6.069} & 6.047 & 5.912 & 5.825 & 5.747 & 5.774 & 5.852 & 5.582 & 5.523 \\
Precision $\uparrow$ & \textbf{0.928} & 0.924 & 0.904 & 0.879 & 0.879 & 0.883 & 0.895 & 0.853 & 0.844 \\
Accuracy $\uparrow$ & \textbf{0.977} & 0.975 & 0.971 & 0.969 & 0.966 & 0.963 & 0.959 & 0.954 & 0.955 \\
MAE $\downarrow$ & \textbf{0.018} & 0.020 & 0.024 & 0.021 & 0.022 & 0.023 & 0.024 & 0.028 & 0.035 \\
R2 $\uparrow$ & 0.866 & 0.848 & 0.821 & 0.867 & \textbf{0.869} & 0.862 & 0.863 & 0.842 & 0.785 \\
\midrule
$\kappa_{\mathrm{SRME}}$ $\downarrow$ & \textbf{0.170} & 1.771 & 0.210 & 0.317 & 0.687 & 0.294 & 0.574 & 0.412 & N/A \\
RMSD $\downarrow$ & \textbf{0.0608} & 0.0691 & 0.0750 & 0.0639 & 0.0679 & 0.0666 & 0.0733 & 0.0731 & N/A \\
\bottomrule
\end{tabular}
\end{adjustbox}
\end{table*}

\textbf{Radial basis functions} are commonly used to embed interatomic distances\citep{bartok2013representing}. A larger number of basis functions (512 in~\citealt{passaro2023reducing}, as opposed to 10 in \ourmodel's default setting) allows higher-frequency signals to pass through the network. This can lead to the PES being more sensitive to small shifts in the atom positions. In our experiments, using a large number of basis functions breaks conservation for the TM23 tasks, but is able to conserve energy for the MD22 tasks. Using a Bessel radial basis function (as opposed to a Gaussian radial basis in the default setting) does not impact conservation properties in both tasks.

\subsection{Ablation studies}
Many of the architecture choices described above have negligible impact on the test set errors as shown in~\cref{tab:design_test_error}. However, as shown in~\cref{fig:conservation}, they can have a dramatic impact on whether a model is conservative in practice. If a model is found to be conservative, stronger correlations are found between test errors and property prediction tasks (\cref{fig:overview} and \cref{fig:test_error_dev}). 

\section{Experiments}
\label{sec:experiment}

In the previous section, we demonstrated the design of \ourmodel\ results in its ability to be energy-conserving in MD simulations. In this section, we evaluate \ourmodel\ in physical property prediction tasks: (1) materials stability prediction based on geometry optimization; (2) thermal conductivity prediction; and (3) phonon calculation. We also demonstrate the correlation between test energy MAE and physical property prediction tasks for \ourmodel.

\subsection{Matbench Discovery}
\textbf{The Matbench-Discovery benchmark} evaluates a model's ability to predict ground-state (0 K) thermodynamic stability through geometry optimization and energy prediction. It is a widely used benchmark for evaluating ML models in materials discovery. The compliant benchmark only includes models trained on the MPTrj~\citep{jain2013materials, deng2023chgnet} dataset or its subset, which facilitate a fair comparison of model architectures. The F1 score is the primary metric used to rank models. We train an \ourmodel\ with 30M parameters on MPTrj for 60 epochs of direct-force pre-training and 40 epochs of conservative fine-tuning. DeNS~\citep{liao2024generalizing} is used during direct-force pre-training. As shown in \cref{tab:mbd}, \ourmodel-30M-MP achieves an F1 score of $0.831$---the highest among all compliant models. \ourmodel-30M-MP also achieves the lowest root mean square deviation (RMSD) when comparing the relaxed structures to the ground truth DFT reference.

\textbf{The thermal conductivity prediction task} requires accurate modeling of harmonic and anharmonic phonons in materials, which tests the accuracy of second and third order derivatives of the learned PES. The primary metric is the symmetric relative mean error in predicting thermal conductivity ($\kappa_{\mathrm{SRME}}$). We follow the protocol set forth in the Matbench-Discovery benchmark \cite{riebesell2023matbench, pota2024thermal} to predict thermal conductivity $\kappa$. After running a structural relaxation, $\kappa$ is computed using second and third order force constants obtained from phonon calculations using the supercell method. 

As shown in \cref{tab:mbd}, our model achieves a $\kappa_{\mathrm{SRME}}$ of $0.340$ under the default evaluation protocol proposed by \citealt{pota2024thermal}. Notably, our model excels in both the F1 score and $\kappa_{\mathrm{SRME}}$, while all previous models only achieve SOTA performance on one or the other of these metrics. 

\textbf{The non-compliant Matbench-Discovery benchmark} includes models trained on datasets other than MPTrj. \ourmodel-30M-OAM is an \ourmodel~model with 30 million parameters pre-trained on the OMat24~\citep{barroso2024open} dataset then fine-tuned on the subsampled Alexandria (sAlex) dataset~\citep{barroso2024open, schmidt2024improving} and MPTrj dataset. As shown in \cref{tab:mbd_nc}, \ourmodel-30M-OAM achieves the best performance among all non-compliant models with an F1 score of $0.925$, a $\kappa_{\mathrm{SRME}}$ of $0.170$, and an RMSD of $0.0608$, significantly advancing state-of-the-art.

\subsection{MDR phonon benchmark}

\begin{figure*}[t]
\includegraphics[width=\textwidth]{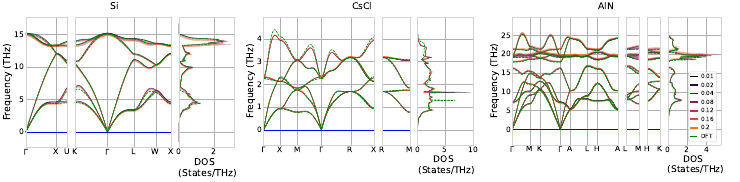}
\caption{
Predicted phonon band structure and density of states (DOS) of Si (diamond structure), CsCl (CsCl structure), AlN (wurtzite structure) using \ourmodel~at different displacement values. DFT baseline is taken from the PBE MDR dataset \cite{loew2024universal} calculated using a displacement of 0.01 \AA.}
\label{fig:phonon_bands}
\end{figure*}

The MDR Phonon benchmark~\citep{loew2024universal} assesses the performance of MLIPs in predicting key phonon properties, including maximum phonon frequency ($\omega_{\text{max}}$), entropy ($S$), free energy ($F$) and heat capacity at constant volume ($C_V$), for around 10,000 materials. The evaluation follows the testing protocol outlined by \citealt{loew2024universal}. \cref{tab:mdr-phonons} shows the resulting MAE of our model and those of several other models\footnote{In addition to our model, we also run the evaluation for GRACE~\citep{bochkarev2024graph}, SevenNet~\citep{park2024scalable}, Orb~\citep{neumann2024orb}, and eqV2-S-DeNS~\citep{liao2023equiformerv2, barroso2024open}, which were not included in the work by \citealt{loew2024universal}.}. \ourmodel\ achieves SOTA results in both compliant and non-compliant categories. 

Our results are consistent with those reported by \citealt{loew2024universal}, showing that conservative MLIPs significantly outperform direct-force models in terms of prediction accuracy when tested using phonon calculations with a displacement of 0.01 \AA. The high error of direct-force models can be largely attributed to high-frequency prediction errors at small displacements \cite{loew2024universal}. Increasing the displacement used in the finite-difference phonon calculations to 0.2 \AA\ can considerably improve prediction accuracy of direct-force models (with caveats). We include a more detailed analysis of the relationship between atom displacement and phonon prediction in \cref{appendix:phonons}.

\begin{table}[t]
    \centering
    \label{tab:mdr-phonons}
    \caption{Summary of model performance on the MDR Phonon benchmark. Metrics include maximum phonon frequency (MAE($\omega_{\mathrm{max}}$), in K), the vibrational entropy (MAE($S$), in J/K/mol), the Helmholtz free energy (MAE($F$), in kJ/mol), and the heat capacity at constant volume (MAE($C_V$), in J/K/mol). \textsuperscript{*}The OMat24 dataset uses a slightly different DFT setting from the DFT setting of the MDR Phonon benchmark.
    \label{tab:mae_models}}
    \begin{adjustbox}{width=\linewidth}
    \begin{tabular}{lcccccc}
    \toprule
    \textit{Compliant models} & MAE($\omega_{\text{max}}$) & MAE($S$) & MAE($F$) & MAE($C_V$) \\
    \midrule
    M3GNet & 98 & 150 & 56 & 22 \\
    CHGNet & 89 & 114 & 45 & 21 \\
    MACE & 61 & 60 & 24 & 13 \\
    GRACE-2L (r6) & 40 & 25 & 9 & 5 \\
    SevenNet-0 & 40 & 48 & 19 & 9 \\
    SevenNet-l3i5 & 26 & 28 & 10 & 5 \\
    \ourmodel-30M-MP & \textbf{21} & \textbf{13} & \textbf{5} & \textbf{4} \\
    \midrule
    \textit{Direct-force models} & & & & \\
    \midrule
    Orb MPTrj [0.01 \AA] & 309 & 476 & 64 & 181 \\
    Orb MPTrj [0.2 \AA] & 61 & 34 & 11 & 8 \\
    eqV2-S-DeNS [0.01 \AA] & 280 & 224 & 54 & 94 \\
    eqV2-S-DeNS [0.2 \AA] & 58 & 26 & 8 & 8 \\
    \midrule
    \textit{Non-compliant models} & & & & \\
    \midrule
    eqV2-M-OAM [0.01 \AA] & 780 & 403 & 241 & 100 \\    
    eqV2-M-OAM [0.2 \AA] & 50 & 25 & 7 & 9 \\    
    MatterSim & 17 & 15 & 5 & 3 \\
    GRACE-2L-OAM & 19 & 14 & 5 & 4 \\
    SevenNet-MF-ompa &  \textbf{15} & \textbf{8} & \textbf{3} & \textbf{3} \\
    \ourmodel-30M-OAM & \textbf{15} & 10 & 4 & \textbf{3} \\
    \bottomrule
    \end{tabular}
    \end{adjustbox}
\end{table}

In physical phonon calculations, we expect the results to converge as the displacement goes to zero. By examining the resulting phonon band structure, we can gain insight into this behavior. \cref{fig:phonon_bands} presents the predicted phonon band structure and density of states for three representative materials using \ourmodel. The predicted phonon bands exhibit convergence as the displacement decreases. In contrast, Figures \cref{fig:phonon_bands_eqV2} and \cref{fig:phonon_bands_eqV2_zero_force} display the phonon bands for the same three materials predicted using eqV2-S-DeNS (direct-forces), which not only fail to demonstrate convergence but also exhibit significant errors, including missing acoustic branches and spurious imaginary frequencies. 

While we find that the OMat-trained models (without sAlex/MPTrj finetuning) may provide a lower error on phonon prediction (7/7/2/2 for $\omega_{\mathrm{max}}$/$S$/$F$/$C_V$ MAE), we refrain from direct comparison due to mismatch in level of theory. We attribute this result to the softening issue of the sAlex and MPTrj dataset~\citep{deng2025systematic, barroso2024open}, which OMat24 addresses. We refer interested readers to \citealt{deng2025systematic} and \citealt{barroso2024open} for a detailed discussion on the softening issue of some DFT datasets and its relation to phonon properties.

\subsection{SPICE-MACE-OFF}

We train and evaluate \ourmodel\ models on the SPICE-MACE-OFF dataset~\citep{kovacs2023mace}, which is built upon the SPICE dataset~\citep{eastman2023spice}. As shown in \cref{tab:spice_mace_off}, \ourmodel\ with 6.5M parameters outperforms MACE-OFF-L (4.7M parameters) and EscAIP (45M parameters, direct-force) on all test-set splits for both energy and force MAE. We also include results for \ourmodel\ with 3.2M parameters, which has inference efficiency similar to MACE-4.7M, while achieving lower test energy/force MAE. More details on the inference efficiency benchmark are included in \cref{appendix:inference}.

\begin{table}[t]
\caption{
Test set MAE for SPICE-MACE-OFF. Energy (\textbf{E}) MAE is in meV/atom. Force (\textbf{F}) MAE is in meV/\AA. \textsuperscript{*}EscAIP-45M is a direct-force model.
\label{tab:spice_mace_off}
}
\begin{adjustbox}{width=\linewidth}
\begin{tabular}{c|cc|cc|cc|cc}
\toprule
 & \multicolumn{2}{c|}{MACE-4.7M} & \multicolumn{2}{c|}{EScAIP-45M\textsuperscript{*}} & \multicolumn{2}{c|}{\ourmodel-3.2M} & \multicolumn{2}{c}{\ourmodel-6.5M} \\
 
 Dataset & \textbf{E} & \textbf{F} & \textbf{E} & \textbf{F} & \textbf{E} & \textbf{F} & \textbf{E} & \textbf{F} \\
\midrule
PubChem & 0.88 & 14.75 & 0.53 & 5.86 & 0.22 & 6.10 & \textbf{0.15} & \textbf{4.21} \\
DES370K M. & 0.59 & 6.58 & 0.41 & 3.48 & 0.17 & 1.85 & \textbf{0.13} & \textbf{1.24} \\
DES370K D. & 0.54 & 6.62 & 0.38 & 2.18 & 0.20 & 2.77 & \textbf{0.15} & \textbf{2.12} \\
Dipeptides & 0.42 & 10.19 & 0.31 & 5.21 & 0.10 & 3.04 & \textbf{0.07} & \textbf{2.00} \\
Sol. AA & 0.98 & 19.43 & 0.61 & 11.52 & 0.30 & 5.76 & \textbf{0.25} & \textbf{3.68} \\
Water & 0.83 & 13.57 & 0.72 & 10.31 & 0.24 & 3.88 & \textbf{0.15} & \textbf{2.50} \\
QMugs & 0.45 & 16.93 & 0.41 & 8.74 & 0.16 & 5.70 & \textbf{0.12} & \textbf{3.78} \\
\bottomrule
\end{tabular}
\end{adjustbox}
\end{table}

\begin{figure}[t]
\includegraphics[width=\linewidth]{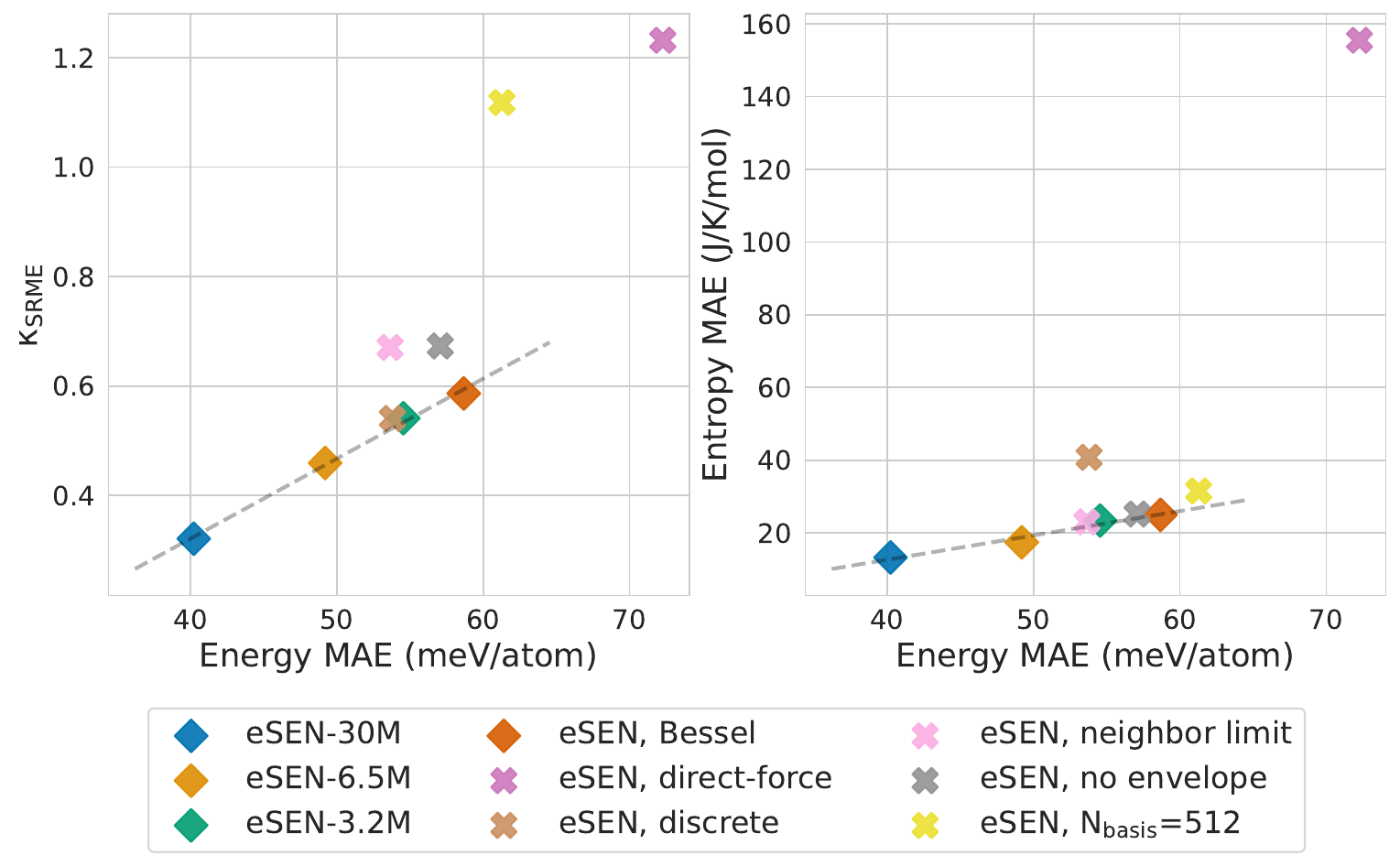}
\caption{ Test error correlation across several property prediction tasks for \ourmodel\ variants. Conservative models are shown as boxes and those found to not conserve as crosses. Note metrics for conservative models have a stronger correlation with test set errors.}
\label{fig:test_error_dev}
\end{figure}

\subsection{Test-set error for model development}

In \cref{fig:overview}, we showed the correlation between test error and physical property prediction tasks for different architectures. \cref{fig:test_error_dev} demonstrates this correlation for different variants of \ourmodel\ (with a 1k-materials subset of the MDR Phonon benchmark for efficiency). In particular, among models that pass the MD energy conservation test, a strong correlation between test error and $\kappa_{\mathrm{SRME}}$/vibrational entropy MAE can be observed. 
We include experimental details about \cref{fig:overview} and \cref{fig:test_error_dev} in \cref{appendix:experimental} and additional results for other phonon properties in \cref{appendix:phonons}.

\section{Related works}

\textbf{MLIP architectures} have made significant progress since their initial proposal~\citep{behler2007generalized}. These architectures are usually symmetry-preserving~\citep{smith2017ani, schutt2017schnet, gilmer2017neural, chmiela2017machine, artrith2017efficient, unke2018reactive, zhang2018end, zubatyuk2019accurate, smith2020ani, kovacs2021linear}, with increasingly expressive atom environment embeddings and message-passing operations~\citep{klicpera2020Directional, gasteiger2021gemnet, schutt2021equivariant, liu2021spherical, unke2021spookynet, chen2022universal, deng2023chgnet, cheng2024cartesian, yin2025alphanet}. Notably, equivariant architectures based on spherical harmonics representations~\citep{thomas2018tensor, tholke2021equivariant, batzner20223, musaelian2022learning, batatia2022mace, passaro2023reducing, liao2023equiformerv2, bochkarev2024graph, park2024scalable, batatia2025design} have shown strong performance on large-scale datasets. Meanwhile, the high computational cost of these architectures has sparked significant interest in scalable architectures that may not respect physical principles such as energy conservation~\citep{langer2024probing, brehmer2024does, hu2021forcenet, yang2024mattersim, qu2024the, neumann2024orb, rhodes2025orb}. These models have demonstrated strong performance in accuracy, scalability, and relaxation tasks~\citep{oc20, riebesell2023matbench}. While their non-physical nature may make them unsuitable for direct usage in some physical property prediction tasks, they may still provide benefit by using the pre-training strategy proposed in this paper and \citealt{bigi2024dark}, distilling them to conservative models~\citep{amin2025towards}, or combining them with a conservative model using multiple-time-step integration~\citep{bigi2024dark}. 

\textbf{MLIPs and physical observables.} While MLIPs continue to improve, it is necessary to evaluate them in realistic tasks that are relevant to scientific discovery. Physical property prediction benchmarks that involve geometry optimization~\citep{riebesell2023matbench, lan2023adsorbml, wander2024cattsunami}, MD simulations~\citep{fu2023forces, kovacs2023mace, moore2024computing, sabanes2024enhancing, eastman2024nutmeg}, vibrational analysis and phonon calculations~\citep{pota2024thermal, loew2024universal, wines2024chips}, and others are increasing in scale with broader applications and wider adoption. Training strategies for learning from physical observables~\citep{wang2020differentiable,  greener2024differentiable, rocken2024predicting, raja2024stability} and the higher-order derivatives of the PES~\citep{fang2024phonon, williams2025hessian} are promising directions to further improve MLIPs for predicting physical properties. 

\section{Discussion}

We identify conservative forces and a smoothly-varying PES as two important properties for MLIPs to consistently perform well in physical property prediction tasks. We offer an analysis of design choices to enhance these two properties. The resulting \ourmodel\ architecture bridges the gap between the test-set error and downstream applications, achieving SOTA performance in force/energy prediction, geometry optimization, phonon calculations, and thermal conductivity prediction. This implies it may be possible to use test error as a proxy metric for evaluating model performance during development, if a model passes energy conservation tests. This can accelerate innovations in MLIPs, since benchmarking physical properties usually requires significant domain knowledge and is usually time-consuming, whereas evaluating test set error is straightforward and efficient. 

\section*{Acknowledgements}

We acknowledge Zachary W.~Ulissi (FAIR at Meta), Aditi S.~Krishnapriyan (UC Berkeley), Samuel M.~Blau (LBNL) and other members of the FAIR Chemistry team for helpful discussions, and Ammar Rizvi (FAIR at Meta) for project support.

\bibliography{paper}

\begin{thebibliography}{94}
\providecommand{\natexlab}[1]{#1}
\providecommand{\url}[1]{\texttt{#1}}
\expandafter\ifx\csname urlstyle\endcsname\relax
  \providecommand{\doi}[1]{doi: #1}\else
  \providecommand{\doi}{doi: \begingroup \urlstyle{rm}\Url}\fi

\bibitem[Ackland et~al.(1997)Ackland, Warren, and Clark]{ackland1997practical}
GJ~Ackland, MC~Warren, and SJ~Clark.
\newblock Practical methods in ab initio lattice dynamics.
\newblock \emph{Journal of Physics: Condensed Matter}, 9\penalty0 (37):\penalty0 7861, 1997.

\bibitem[Amin et~al.(2025)Amin, Raja, and Krishnapriyan]{amin2025towards}
Ishan Amin, Sanjeev Raja, and Aditi Krishnapriyan.
\newblock Towards fast, specialized machine learning force fields: Distilling foundation models via energy hessians.
\newblock \emph{arXiv preprint arXiv:2501.09009}, 2025.

\bibitem[Artrith et~al.(2017)Artrith, Urban, and Ceder]{artrith2017efficient}
Nongnuch Artrith, Alexander Urban, and Gerbrand Ceder.
\newblock Efficient and accurate machine-learning interpolation of atomic energies in compositions with many species.
\newblock \emph{Physical Review B}, 96\penalty0 (1):\penalty0 014112, 2017.

\bibitem[Ba(2016)]{ba2016layer}
Jimmy~Lei Ba.
\newblock Layer normalization.
\newblock \emph{arXiv preprint arXiv:1607.06450}, 2016.

\bibitem[Barroso-Luque et~al.(2024)Barroso-Luque, Shuaibi, Fu, Wood, Dzamba, Gao, Rizvi, Zitnick, and Ulissi]{barroso2024open}
Luis Barroso-Luque, Muhammed Shuaibi, Xiang Fu, Brandon~M Wood, Misko Dzamba, Meng Gao, Ammar Rizvi, C~Lawrence Zitnick, and Zachary~W Ulissi.
\newblock Open materials 2024 (omat24) inorganic materials dataset and models.
\newblock \emph{arXiv preprint arXiv:2410.12771}, 2024.

\bibitem[Bartel(2022)]{bartel2022review}
Christopher~J Bartel.
\newblock Review of computational approaches to predict the thermodynamic stability of inorganic solids.
\newblock \emph{Journal of Materials Science}, 57\penalty0 (23):\penalty0 10475--10498, 2022.

\bibitem[Bart{\'o}k et~al.(2013)Bart{\'o}k, Kondor, and Cs{\'a}nyi]{bartok2013representing}
Albert~P Bart{\'o}k, Risi Kondor, and G{\'a}bor Cs{\'a}nyi.
\newblock On representing chemical environments.
\newblock \emph{Physical Review B—Condensed Matter and Materials Physics}, 87\penalty0 (18):\penalty0 184115, 2013.

\bibitem[Batatia et~al.(2022)Batatia, Kovacs, Simm, Ortner, and Csanyi]{batatia2022mace}
Ilyes Batatia, David~Peter Kovacs, Gregor N.~C. Simm, Christoph Ortner, and Gabor Csanyi.
\newblock {MACE}: Higher order equivariant message passing neural networks for fast and accurate force fields.
\newblock In Alice~H. Oh, Alekh Agarwal, Danielle Belgrave, and Kyunghyun Cho, editors, \emph{Advances in Neural Information Processing Systems}, 2022.
\newblock \url{https://openreview.net/forum?id=YPpSngE-ZU}.

\bibitem[Batatia et~al.(2023)Batatia, Benner, Chiang, Elena, Kov{\'a}cs, Riebesell, Advincula, Asta, Avaylon, Baldwin, et~al.]{batatia2023foundation}
Ilyes Batatia, Philipp Benner, Yuan Chiang, Alin~M Elena, D{\'a}vid~P Kov{\'a}cs, Janosh Riebesell, Xavier~R Advincula, Mark Asta, Matthew Avaylon, William~J Baldwin, et~al.
\newblock A foundation model for atomistic materials chemistry.
\newblock \emph{arXiv preprint arXiv:2401.00096}, 2023.

\bibitem[Batatia et~al.(2025)Batatia, Batzner, Kov{\'a}cs, Musaelian, Simm, Drautz, Ortner, Kozinsky, and Cs{\'a}nyi]{batatia2025design}
Ilyes Batatia, Simon Batzner, D{\'a}vid~P{\'e}ter Kov{\'a}cs, Albert Musaelian, Gregor~NC Simm, Ralf Drautz, Christoph Ortner, Boris Kozinsky, and G{\'a}bor Cs{\'a}nyi.
\newblock The design space of e (3)-equivariant atom-centred interatomic potentials.
\newblock \emph{Nature Machine Intelligence}, pages 1--12, 2025.

\bibitem[Batzner et~al.(2022)Batzner, Musaelian, Sun, Geiger, Mailoa, Kornbluth, Molinari, Smidt, and Kozinsky]{batzner20223}
Simon Batzner, Albert Musaelian, Lixin Sun, Mario Geiger, Jonathan~P Mailoa, Mordechai Kornbluth, Nicola Molinari, Tess~E Smidt, and Boris Kozinsky.
\newblock E (3)-equivariant graph neural networks for data-efficient and accurate interatomic potentials.
\newblock \emph{Nature communications}, 13\penalty0 (1):\penalty0 1--11, 2022.

\bibitem[Behler and Parrinello(2007)]{behler2007generalized}
J{\"o}rg Behler and Michele Parrinello.
\newblock Generalized neural-network representation of high-dimensional potential-energy surfaces.
\newblock \emph{Physical review letters}, 98\penalty0 (14):\penalty0 146401, 2007.

\bibitem[Bigi et~al.(2024)Bigi, Langer, and Ceriotti]{bigi2024dark}
Filippo Bigi, Marcel Langer, and Michele Ceriotti.
\newblock The dark side of the forces: assessing non-conservative force models for atomistic machine learning.
\newblock \emph{arXiv preprint arXiv:2412.11569}, 2024.

\bibitem[Bochkarev et~al.(2024)Bochkarev, Lysogorskiy, and Drautz]{bochkarev2024graph}
Anton Bochkarev, Yury Lysogorskiy, and Ralf Drautz.
\newblock Graph atomic cluster expansion for semilocal interactions beyond equivariant message passing.
\newblock \emph{Phys. Rev. X}, 14:\penalty0 021036, Jun 2024.
\newblock \doi{10.1103/PhysRevX.14.021036}.
\newblock \url{https://link.aps.org/doi/10.1103/PhysRevX.14.021036}.

\bibitem[Brehmer et~al.(2024)Brehmer, Behrends, de~Haan, and Cohen]{brehmer2024does}
Johann Brehmer, S{\"o}nke Behrends, Pim de~Haan, and Taco Cohen.
\newblock Does equivariance matter at scale?
\newblock \emph{arXiv preprint arXiv:2410.23179}, 2024.

\bibitem[Chanussot et~al.(2021)Chanussot, Das, Goyal, Lavril, Shuaibi, Riviere, Tran, Heras-Domingo, Ho, Hu, Palizhati, Sriram, Wood, Yoon, Parikh, Zitnick, and Ulissi]{oc20}
Lowik Chanussot, Abhishek Das, Siddharth Goyal, Thibaut Lavril, Muhammed Shuaibi, Morgane Riviere, Kevin Tran, Javier Heras-Domingo, Caleb Ho, Weihua Hu, Aini Palizhati, Anuroop Sriram, Brandon Wood, Junwoong Yoon, Devi Parikh, C.~Lawrence Zitnick, and Zachary Ulissi.
\newblock Open catalyst 2020 (oc20) dataset and community challenges.
\newblock \emph{ACS Catalysis}, 11\penalty0 (10):\penalty0 6059--6072, 2021.
\newblock \doi{10.1021/acscatal.0c04525}.
\newblock \url{https://doi.org/10.1021/acscatal.0c04525}.

\bibitem[Chen and Ong(2022)]{chen2022universal}
Chi Chen and Shyue~Ping Ong.
\newblock A universal graph deep learning interatomic potential for the periodic table.
\newblock \emph{Nature Computational Science}, 2\penalty0 (11):\penalty0 718--728, 2022.

\bibitem[Cheng(2024)]{cheng2024cartesian}
Bingqing Cheng.
\newblock Cartesian atomic cluster expansion for machine learning interatomic potentials.
\newblock \emph{npj Computational Materials}, 10\penalty0 (1):\penalty0 157, 2024.

\bibitem[Chmiela et~al.(2017)Chmiela, Tkatchenko, Sauceda, Poltavsky, Sch{\"u}tt, and M{\"u}ller]{chmiela2017machine}
Stefan Chmiela, Alexandre Tkatchenko, Huziel~E Sauceda, Igor Poltavsky, Kristof~T Sch{\"u}tt, and Klaus-Robert M{\"u}ller.
\newblock Machine learning of accurate energy-conserving molecular force fields.
\newblock \emph{Science advances}, 3\penalty0 (5):\penalty0 e1603015, 2017.

\bibitem[Chmiela et~al.(2023)Chmiela, Vassilev-Galindo, Unke, Kabylda, Sauceda, Tkatchenko, and M{\"u}ller]{chmiela2023accurate}
Stefan Chmiela, Valentin Vassilev-Galindo, Oliver~T Unke, Adil Kabylda, Huziel~E Sauceda, Alexandre Tkatchenko, and Klaus-Robert M{\"u}ller.
\newblock Accurate global machine learning force fields for molecules with hundreds of atoms.
\newblock \emph{Science Advances}, 9\penalty0 (2):\penalty0 eadf0873, 2023.

\bibitem[Cohen and Welling(2016)]{cohen2016group}
Taco Cohen and Max Welling.
\newblock Group equivariant convolutional networks.
\newblock In \emph{International conference on machine learning}, pages 2990--2999. PMLR, 2016.

\bibitem[Deng et~al.(2023)Deng, Zhong, Jun, Riebesell, Han, Bartel, and Ceder]{deng2023chgnet}
Bowen Deng, Peichen Zhong, KyuJung Jun, Janosh Riebesell, Kevin Han, Christopher~J Bartel, and Gerbrand Ceder.
\newblock Chgnet as a pretrained universal neural network potential for charge-informed atomistic modelling.
\newblock \emph{Nature Machine Intelligence}, 5\penalty0 (9):\penalty0 1031--1041, 2023.

\bibitem[Deng et~al.(2025)Deng, Choi, Zhong, Riebesell, Anand, Li, Jun, Persson, and Ceder]{deng2025systematic}
Bowen Deng, Yunyeong Choi, Peichen Zhong, Janosh Riebesell, Shashwat Anand, Zhuohan Li, KyuJung Jun, Kristin~A Persson, and Gerbrand Ceder.
\newblock Systematic softening in universal machine learning interatomic potentials.
\newblock \emph{npj Computational Materials}, 11\penalty0 (1):\penalty0 1--9, 2025.

\bibitem[Eastman et~al.(2023)Eastman, Behara, Dotson, Galvelis, Herr, Horton, Mao, Chodera, Pritchard, Wang, et~al.]{eastman2023spice}
Peter Eastman, Pavan~Kumar Behara, David~L Dotson, Raimondas Galvelis, John~E Herr, Josh~T Horton, Yuezhi Mao, John~D Chodera, Benjamin~P Pritchard, Yuanqing Wang, et~al.
\newblock Spice, a dataset of drug-like molecules and peptides for training machine learning potentials.
\newblock \emph{Scientific Data}, 10\penalty0 (1):\penalty0 11, 2023.

\bibitem[Eastman et~al.(2024)Eastman, Pritchard, Chodera, and Markland]{eastman2024nutmeg}
Peter Eastman, Benjamin~P Pritchard, John~D Chodera, and Thomas~E Markland.
\newblock Nutmeg and spice: models and data for biomolecular machine learning.
\newblock \emph{Journal of chemical theory and computation}, 20\penalty0 (19):\penalty0 8583--8593, 2024.

\bibitem[Fang et~al.(2024)Fang, Geiger, Checkelsky, and Smidt]{fang2024phonon}
Shiang Fang, Mario Geiger, Joseph~G Checkelsky, and Tess Smidt.
\newblock Phonon predictions with e (3)-equivariant graph neural networks.
\newblock \emph{arXiv preprint arXiv:2403.11347}, 2024.

\bibitem[Fu et~al.(2023)Fu, Wu, Wang, Xie, Keten, Gomez-Bombarelli, and Jaakkola]{fu2023forces}
Xiang Fu, Zhenghao Wu, Wujie Wang, Tian Xie, Sinan Keten, Rafael Gomez-Bombarelli, and Tommi~S. Jaakkola.
\newblock Forces are not enough: Benchmark and critical evaluation for machine learning force fields with molecular simulations.
\newblock \emph{Transactions on Machine Learning Research}, 2023.
\newblock ISSN 2835-8856.
\newblock \url{https://openreview.net/forum?id=A8pqQipwkt}.
\newblock Survey Certification.

\bibitem[Fultz(2010)]{fultz2010vibrational}
Brent Fultz.
\newblock Vibrational thermodynamics of materials.
\newblock \emph{Progress in Materials Science}, 55\penalty0 (4):\penalty0 247--352, 2010.

\bibitem[Ganose et~al.(2021)Ganose, Park, Faghaninia, Woods-Robinson, Persson, and Jain]{ganose2021efficient}
Alex~M Ganose, Junsoo Park, Alireza Faghaninia, Rachel Woods-Robinson, Kristin~A Persson, and Anubhav Jain.
\newblock Efficient calculation of carrier scattering rates from first principles.
\newblock \emph{Nature communications}, 12\penalty0 (1):\penalty0 2222, 2021.

\bibitem[Gasteiger et~al.(2020{\natexlab{a}})Gasteiger, Gro{\ss}, and G{\"u}nnemann]{gasteiger2020directional}
Johannes Gasteiger, Janek Gro{\ss}, and Stephan G{\"u}nnemann.
\newblock Directional message passing for molecular graphs.
\newblock \emph{arXiv preprint arXiv:2003.03123}, 2020{\natexlab{a}}.

\bibitem[Gasteiger et~al.(2020{\natexlab{b}})Gasteiger, Gro{\ss}, and G{\"u}nnemann]{klicpera2020Directional}
Johannes Gasteiger, Janek Gro{\ss}, and Stephan G{\"u}nnemann.
\newblock Directional message passing for molecular graphs.
\newblock In \emph{International Conference on Learning Representations}, 2020{\natexlab{b}}.

\bibitem[Gasteiger et~al.(2021)Gasteiger, Becker, and G{\"u}nnemann]{gasteiger2021gemnet}
Johannes Gasteiger, Florian Becker, and Stephan G{\"u}nnemann.
\newblock Gemnet: Universal directional graph neural networks for molecules.
\newblock \emph{Advances in Neural Information Processing Systems}, 34:\penalty0 6790--6802, 2021.

\bibitem[Geiger and Smidt(2022)]{geiger2022e3nn}
Mario Geiger and Tess Smidt.
\newblock e3nn: Euclidean neural networks.
\newblock \emph{arXiv preprint arXiv:2207.09453}, 2022.

\bibitem[Gilmer et~al.(2017)Gilmer, Schoenholz, Riley, Vinyals, and Dahl]{gilmer2017neural}
Justin Gilmer, Samuel~S Schoenholz, Patrick~F Riley, Oriol Vinyals, and George~E Dahl.
\newblock Neural message passing for quantum chemistry.
\newblock In \emph{International conference on machine learning}, pages 1263--1272. PMLR, 2017.

\bibitem[Greener(2024)]{greener2024differentiable}
Joe~G Greener.
\newblock Differentiable simulation to develop molecular dynamics force fields for disordered proteins.
\newblock \emph{Chemical Science}, 15\penalty0 (13):\penalty0 4897--4909, 2024.

\bibitem[Hairer et~al.(2003)Hairer, Lubich, and Wanner]{hairer2003geometric}
Ernst Hairer, Christian Lubich, and Gerhard Wanner.
\newblock Geometric numerical integration illustrated by the st{\"o}rmer--verlet method.
\newblock \emph{Acta numerica}, 12:\penalty0 399--450, 2003.

\bibitem[Hairer et~al.(2006)Hairer, Hochbruck, Iserles, and Lubich]{hairer2006geometric}
Ernst Hairer, Marlis Hochbruck, Arieh Iserles, and Christian Lubich.
\newblock Geometric numerical integration.
\newblock \emph{Oberwolfach Reports}, 3\penalty0 (1):\penalty0 805--882, 2006.

\bibitem[Hu et~al.(2021)Hu, Shuaibi, Das, Goyal, Sriram, Leskovec, Parikh, and Zitnick]{hu2021forcenet}
Weihua Hu, Muhammed Shuaibi, Abhishek Das, Siddharth Goyal, Anuroop Sriram, Jure Leskovec, Devi Parikh, and C~Lawrence Zitnick.
\newblock Forcenet: A graph neural network for large-scale quantum calculations.
\newblock \emph{arXiv preprint arXiv:2103.01436}, 2021.

\bibitem[Jain et~al.(2013)Jain, Ong, Hautier, Chen, Richards, Dacek, Cholia, Gunter, Skinner, Ceder, et~al.]{jain2013materials}
A~Jain, SP~Ong, G~Hautier, W~Chen, WD~Richards, S~Dacek, S~Cholia, D~Gunter, D~Skinner, G~Ceder, et~al.
\newblock The materials project: a materials genome approach to accelerating materials innovation. apl mater 1: 011002, 2013.

\bibitem[Kov{\'a}cs et~al.(2021)Kov{\'a}cs, Oord, Kucera, Allen, Cole, Ortner, and Cs{\'a}nyi]{kovacs2021linear}
D{\'a}vid~P{\'e}ter Kov{\'a}cs, Cas van~der Oord, Jiri Kucera, Alice~EA Allen, Daniel~J Cole, Christoph Ortner, and G{\'a}bor Cs{\'a}nyi.
\newblock Linear atomic cluster expansion force fields for organic molecules: beyond rmse.
\newblock \emph{Journal of chemical theory and computation}, 17\penalty0 (12):\penalty0 7696--7711, 2021.

\bibitem[Kov{\'a}cs et~al.(2023)Kov{\'a}cs, Moore, Browning, Batatia, Horton, Kapil, Witt, Magd{\u{a}}u, Cole, and Cs{\'a}nyi]{kovacs2023mace}
D{\'a}vid~P{\'e}ter Kov{\'a}cs, J~Harry Moore, Nicholas~J Browning, Ilyes Batatia, Joshua~T Horton, Venkat Kapil, William~C Witt, Ioan-Bogdan Magd{\u{a}}u, Daniel~J Cole, and G{\'a}bor Cs{\'a}nyi.
\newblock Mace-off23: Transferable machine learning force fields for organic molecules.
\newblock \emph{arXiv preprint arXiv:2312.15211}, 2023.

\bibitem[Lan et~al.(2023)Lan, Palizhati, Shuaibi, Wood, Wander, Das, Uyttendaele, Zitnick, and Ulissi]{lan2023adsorbml}
Janice Lan, Aini Palizhati, Muhammed Shuaibi, Brandon~M Wood, Brook Wander, Abhishek Das, Matt Uyttendaele, C~Lawrence Zitnick, and Zachary~W Ulissi.
\newblock Adsorbml: a leap in efficiency for adsorption energy calculations using generalizable machine learning potentials.
\newblock \emph{npj Computational Materials}, 9\penalty0 (1):\penalty0 172, 2023.

\bibitem[Langer et~al.(2024)Langer, Pozdnyakov, and Ceriotti]{langer2024probing}
Marcel~F Langer, Sergey~N Pozdnyakov, and Michele Ceriotti.
\newblock Probing the effects of broken symmetries in machine learning.
\newblock \emph{Machine Learning: Science and Technology}, 5\penalty0 (4):\penalty0 04LT01, 2024.

\bibitem[Larsen et~al.(2017)Larsen, Mortensen, Blomqvist, Castelli, Christensen, Dułak, Friis, Groves, Hammer, Hargus, Hermes, Jennings, Jensen, Kermode, Kitchin, Kolsbjerg, Kubal, Kaasbjerg, Lysgaard, Maronsson, Maxson, Olsen, Pastewka, Peterson, Rostgaard, Schiøtz, Schütt, Strange, Thygesen, Vegge, Vilhelmsen, Walter, Zeng, and Jacobsen]{ase}
Ask~Hjorth Larsen, Jens~Jørgen Mortensen, Jakob Blomqvist, Ivano~E Castelli, Rune Christensen, Marcin Dułak, Jesper Friis, Michael~N Groves, Bjørk Hammer, Cory Hargus, Eric~D Hermes, Paul~C Jennings, Peter~Bjerre Jensen, James Kermode, John~R Kitchin, Esben~Leonhard Kolsbjerg, Joseph Kubal, Kristen Kaasbjerg, Steen Lysgaard, Jón~Bergmann Maronsson, Tristan Maxson, Thomas Olsen, Lars Pastewka, Andrew Peterson, Carsten Rostgaard, Jakob Schiøtz, Ole Schütt, Mikkel Strange, Kristian~S Thygesen, Tejs Vegge, Lasse Vilhelmsen, Michael Walter, Zhenhua Zeng, and Karsten~W Jacobsen.
\newblock The atomic simulation environment—a python library for working with atoms.
\newblock \emph{Journal of Physics: Condensed Matter}, 29\penalty0 (27):\penalty0 273002, 2017.
\newblock \url{http://stacks.iop.org/0953-8984/29/i=27/a=273002}.

\bibitem[Liao and Smidt(2022)]{liao2022equiformer}
Yi-Lun Liao and Tess Smidt.
\newblock Equiformer: Equivariant graph attention transformer for 3d atomistic graphs.
\newblock \emph{arXiv preprint arXiv:2206.11990}, 2022.

\bibitem[Liao et~al.(2023)Liao, Wood, Das, and Smidt]{liao2023equiformerv2}
Yi-Lun Liao, Brandon Wood, Abhishek Das, and Tess Smidt.
\newblock Equiformerv2: Improved equivariant transformer for scaling to higher-degree representations.
\newblock \emph{arXiv preprint arXiv:2306.12059}, 2023.

\bibitem[Liao et~al.(2024)Liao, Smidt, and Das]{liao2024generalizing}
Yi-Lun Liao, Tess Smidt, and Abhishek Das.
\newblock Generalizing denoising to non-equilibrium structures improves equivariant force fields.
\newblock \emph{arXiv preprint arXiv:2403.09549}, 2024.

\bibitem[Liu et~al.(2021)Liu, Wang, Liu, Lin, Zhang, Oztekin, and Ji]{liu2021spherical}
Yi~Liu, Limei Wang, Meng Liu, Yuchao Lin, Xuan Zhang, Bora Oztekin, and Shuiwang Ji.
\newblock Spherical message passing for 3d molecular graphs.
\newblock In \emph{International Conference on Learning Representations}, 2021.

\bibitem[Loew et~al.(2024)Loew, Sun, Wang, Botti, and Marques]{loew2024universal}
Antoine Loew, Dewen Sun, Hai-Chen Wang, Silvana Botti, and Miguel~AL Marques.
\newblock Universal machine learning interatomic potentials are ready for phonons.
\newblock \emph{arXiv preprint arXiv:2412.16551}, 2024.

\bibitem[Merchant et~al.(2023)Merchant, Batzner, Schoenholz, Aykol, Cheon, and Cubuk]{merchant2023scaling}
Amil Merchant, Simon Batzner, Samuel~S Schoenholz, Muratahan Aykol, Gowoon Cheon, and Ekin~Dogus Cubuk.
\newblock Scaling deep learning for materials discovery.
\newblock \emph{Nature}, 624\penalty0 (7990):\penalty0 80--85, 2023.

\bibitem[Moore et~al.(2024)Moore, Cole, and Csanyi]{moore2024computing}
J~Harry Moore, Daniel~J Cole, and Gabor Csanyi.
\newblock Computing hydration free energies of small molecules with first principles accuracy.
\newblock \emph{arXiv preprint arXiv:2405.18171}, 2024.

\bibitem[Musaelian et~al.(2022)Musaelian, Batzner, Johansson, Sun, Owen, Kornbluth, and Kozinsky]{musaelian2022learning}
Albert Musaelian, Simon Batzner, Anders Johansson, Lixin Sun, Cameron~J Owen, Mordechai Kornbluth, and Boris Kozinsky.
\newblock Learning local equivariant representations for large-scale atomistic dynamics.
\newblock \emph{arXiv preprint arXiv:2204.05249}, 2022.

\bibitem[Neumann et~al.(2024)Neumann, Gin, Rhodes, Bennett, Li, Choubisa, Hussey, and Godwin]{neumann2024orb}
Mark Neumann, James Gin, Benjamin Rhodes, Steven Bennett, Zhiyi Li, Hitarth Choubisa, Arthur Hussey, and Jonathan Godwin.
\newblock Orb: A fast, scalable neural network potential.
\newblock \emph{arXiv preprint arXiv:2410.22570}, 2024.

\bibitem[Nickolls et~al.(2008)Nickolls, Buck, Garland, and Skadron]{nickolls2008scalable}
John Nickolls, Ian Buck, Michael Garland, and Kevin Skadron.
\newblock Scalable parallel programming with cuda: Is cuda the parallel programming model that application developers have been waiting for?
\newblock \emph{Queue}, 6\penalty0 (2):\penalty0 40--53, 2008.

\bibitem[Oppenheimer(1927)]{oppenheimer1927quantentheorie}
MBJR Oppenheimer.
\newblock Zur quantentheorie der molekeln [on the quantum theory of molecules].
\newblock \emph{Annalen der Physik}, 389\penalty0 (20):\penalty0 457--484, 1927.

\bibitem[Owen et~al.(2024)Owen, Torrisi, Xie, Batzner, Bystrom, Coulter, Musaelian, Sun, and Kozinsky]{owen2024complexity}
Cameron~J Owen, Steven~B Torrisi, Yu~Xie, Simon Batzner, Kyle Bystrom, Jennifer Coulter, Albert Musaelian, Lixin Sun, and Boris Kozinsky.
\newblock Complexity of many-body interactions in transition metals via machine-learned force fields from the tm23 data set.
\newblock \emph{npj Computational Materials}, 10\penalty0 (1):\penalty0 92, 2024.

\bibitem[Park et~al.(2024)Park, Kim, Hwang, and Han]{park2024scalable}
Yutack Park, Jaesun Kim, Seungwoo Hwang, and Seungwu Han.
\newblock Scalable parallel algorithm for graph neural network interatomic potentials in molecular dynamics simulations.
\newblock \emph{J. Chem. Theory Comput.}, 20\penalty0 (11):\penalty0 4857--4868, 2024.
\newblock \doi{10.1021/acs.jctc.4c00190}.

\bibitem[Parr et~al.(1979)Parr, Gadre, and Bartolotti]{parr1979local}
Robert~G Parr, Shridhar~R Gadre, and Libero~J Bartolotti.
\newblock Local density functional theory of atoms and molecules.
\newblock \emph{Proceedings of the National Academy of Sciences}, 76\penalty0 (6):\penalty0 2522--2526, 1979.

\bibitem[Passaro and Zitnick(2023)]{passaro2023reducing}
Saro Passaro and C~Lawrence Zitnick.
\newblock Reducing so (3) convolutions to so (2) for efficient equivariant gnns.
\newblock In \emph{International Conference on Machine Learning}, pages 27420--27438. Proceedings of Machine Learning Research, 2023.

\bibitem[Paszke et~al.(2019)Paszke, Gross, Massa, Lerer, Bradbury, Chanan, Killeen, Lin, Gimelshein, Antiga, et~al.]{paszke2019pytorch}
Adam Paszke, Sam Gross, Francisco Massa, Adam Lerer, James Bradbury, Gregory Chanan, Trevor Killeen, Zeming Lin, Natalia Gimelshein, Luca Antiga, et~al.
\newblock Pytorch: An imperative style, high-performance deep learning library.
\newblock \emph{Advances in neural information processing systems}, 32, 2019.

\bibitem[P{\'o}ta et~al.(2024)P{\'o}ta, Ahlawat, Cs{\'a}nyi, and Simoncelli]{pota2024thermal}
Bal{\'a}zs P{\'o}ta, Paramvir Ahlawat, G{\'a}bor Cs{\'a}nyi, and Michele Simoncelli.
\newblock Thermal conductivity predictions with foundation atomistic models.
\newblock \emph{arXiv preprint arXiv:2408.00755}, 2024.

\bibitem[Qu and Krishnapriyan(2024)]{qu2024the}
Eric Qu and Aditi~S. Krishnapriyan.
\newblock The importance of being scalable: Improving the speed and accuracy of neural network interatomic potentials across chemical domains.
\newblock In \emph{The Thirty-eighth Annual Conference on Neural Information Processing Systems}, 2024.
\newblock \url{https://openreview.net/forum?id=Y4mBaZu4vy}.

\bibitem[Raja et~al.(2024)Raja, Amin, Pedregosa, and Krishnapriyan]{raja2024stability}
Sanjeev Raja, Ishan Amin, Fabian Pedregosa, and Aditi~S Krishnapriyan.
\newblock Stability-aware training of neural network interatomic potentials with differentiable boltzmann estimators.
\newblock \emph{arXiv preprint arXiv:2402.13984}, 2024.

\bibitem[Razeghi(2002)]{razeghi2002thermal}
Manijeh Razeghi.
\newblock Thermal {{Properties}} of {{Crystals}}.
\newblock In Manijeh Razeghi, editor, \emph{Fundamentals of {{Solid State Engineering}}}, pages 197--220. Springer US, 2002.
\newblock \doi{10.1007/0-306-47567-7_7}.

\bibitem[Rhodes et~al.(2025)Rhodes, Vandenhaute, {\v{S}}imkus, Gin, Godwin, Duignan, and Neumann]{rhodes2025orb}
Benjamin Rhodes, Sander Vandenhaute, Vaidotas {\v{S}}imkus, James Gin, Jonathan Godwin, Tim Duignan, and Mark Neumann.
\newblock Orb-v3: atomistic simulation at scale.
\newblock \emph{arXiv preprint arXiv:2504.06231}, 2025.

\bibitem[Riebesell et~al.(2023)Riebesell, Goodall, Jain, Benner, Persson, and Lee]{riebesell2023matbench}
Janosh Riebesell, Rhys~EA Goodall, Anubhav Jain, Philipp Benner, Kristin~A Persson, and Alpha~A Lee.
\newblock Matbench discovery--an evaluation framework for machine learning crystal stability prediction.
\newblock \emph{arXiv preprint arXiv:2308.14920}, 2023.

\bibitem[R{\"o}cken et~al.(2024)R{\"o}cken, Burnet, and Zavadlav]{rocken2024predicting}
Sebastien R{\"o}cken, Anton~F Burnet, and Julija Zavadlav.
\newblock Predicting solvation free energies with an implicit solvent machine learning potential.
\newblock \emph{arXiv preprint arXiv:2406.00183}, 2024.

\bibitem[Sabanes~Zariquiey et~al.(2024)Sabanes~Zariquiey, Galvelis, Gallicchio, Chodera, Markland, and De~Fabritiis]{sabanes2024enhancing}
Francesc Sabanes~Zariquiey, Raimondas Galvelis, Emilio Gallicchio, John~D Chodera, Thomas~E Markland, and Gianni De~Fabritiis.
\newblock Enhancing protein--ligand binding affinity predictions using neural network potentials.
\newblock \emph{Journal of Chemical Information and Modeling}, 64\penalty0 (5):\penalty0 1481--1485, 2024.

\bibitem[Schmidt et~al.(2024)Schmidt, Cerqueira, Romero, Loew, J{\"a}ger, Wang, Botti, and Marques]{schmidt2024improving}
Jonathan Schmidt, Tiago~FT Cerqueira, Aldo~H Romero, Antoine Loew, Fabian J{\"a}ger, Hai-Chen Wang, Silvana Botti, and Miguel~AL Marques.
\newblock Improving machine-learning models in materials science through large datasets.
\newblock \emph{Materials Today Physics}, 48:\penalty0 101560, 2024.

\bibitem[Sch{\"u}tt et~al.(2017)Sch{\"u}tt, Kindermans, Sauceda~Felix, Chmiela, Tkatchenko, and M{\"u}ller]{schutt2017schnet}
Kristof Sch{\"u}tt, Pieter-Jan Kindermans, Huziel~Enoc Sauceda~Felix, Stefan Chmiela, Alexandre Tkatchenko, and Klaus-Robert M{\"u}ller.
\newblock Schnet: A continuous-filter convolutional neural network for modeling quantum interactions.
\newblock \emph{Advances in neural information processing systems}, 30, 2017.

\bibitem[Sch{\"u}tt et~al.(2021)Sch{\"u}tt, Unke, and Gastegger]{schutt2021equivariant}
Kristof Sch{\"u}tt, Oliver Unke, and Michael Gastegger.
\newblock Equivariant message passing for the prediction of tensorial properties and molecular spectra.
\newblock In \emph{International Conference on Machine Learning}, pages 9377--9388. PMLR, 2021.

\bibitem[Simoncelli et~al.(2022)Simoncelli, Marzari, and Mauri]{simoncelli2022wigner}
Michele Simoncelli, Nicola Marzari, and Francesco Mauri.
\newblock Wigner formulation of thermal transport in solids.
\newblock \emph{Physical Review X}, 12\penalty0 (4):\penalty0 041011, 2022.

\bibitem[Smith et~al.(2017)Smith, Isayev, and Roitberg]{smith2017ani}
Justin~S Smith, Olexandr Isayev, and Adrian~E Roitberg.
\newblock Ani-1: an extensible neural network potential with dft accuracy at force field computational cost.
\newblock \emph{Chemical science}, 8\penalty0 (4):\penalty0 3192--3203, 2017.

\bibitem[Smith et~al.(2020)Smith, Zubatyuk, Nebgen, Lubbers, Barros, Roitberg, Isayev, and Tretiak]{smith2020ani}
Justin~S Smith, Roman Zubatyuk, Benjamin Nebgen, Nicholas Lubbers, Kipton Barros, Adrian~E Roitberg, Olexandr Isayev, and Sergei Tretiak.
\newblock The ani-1ccx and ani-1x data sets, coupled-cluster and density functional theory properties for molecules.
\newblock \emph{Scientific data}, 7\penalty0 (1):\penalty0 134, 2020.

\bibitem[Th{\"o}lke and De~Fabritiis(2021)]{tholke2021equivariant}
Philipp Th{\"o}lke and Gianni De~Fabritiis.
\newblock Equivariant transformers for neural network based molecular potentials.
\newblock In \emph{International Conference on Learning Representations}, 2021.

\bibitem[Thomas et~al.(2018)Thomas, Smidt, Kearnes, Yang, Li, Kohlhoff, and Riley]{thomas2018tensor}
Nathaniel Thomas, Tess Smidt, Steven Kearnes, Lusann Yang, Li~Li, Kai Kohlhoff, and Patrick Riley.
\newblock Tensor field networks: Rotation-and translation-equivariant neural networks for 3d point clouds.
\newblock \emph{arXiv preprint arXiv:1802.08219}, 2018.

\bibitem[Togo et~al.(2015)Togo, Chaput, and Tanaka]{togo2015phono3py}
Atsushi Togo, Laurent Chaput, and Isao Tanaka.
\newblock Distributions of phonon lifetimes in brillouin zones.
\newblock \emph{Phys. Rev. B}, 91:\penalty0 094306, Mar 2015.
\newblock \doi{10.1103/PhysRevB.91.094306}.

\bibitem[Togo et~al.(2023)Togo, Chaput, Tadano, and Tanaka]{togo2023phonopy}
Atsushi Togo, Laurent Chaput, Terumasa Tadano, and Isao Tanaka.
\newblock Implementation strategies in phonopy and phono3py.
\newblock \emph{J. Phys. Condens. Matter}, 35\penalty0 (35):\penalty0 353001, 2023.
\newblock \doi{10.1088/1361-648X/acd831}.

\bibitem[Tran et~al.(2023)Tran, Lan, Shuaibi, Wood, Goyal, Das, Heras-Domingo, Kolluru, Rizvi, Shoghi, et~al.]{oc22}
Richard Tran, Janice Lan, Muhammed Shuaibi, Brandon~M Wood, Siddharth Goyal, Abhishek Das, Javier Heras-Domingo, Adeesh Kolluru, Ammar Rizvi, Nima Shoghi, et~al.
\newblock The open catalyst 2022 (oc22) dataset and challenges for oxide electrocatalysts.
\newblock \emph{ACS Catalysis}, 13\penalty0 (5):\penalty0 3066--3084, 2023.

\bibitem[Tuckerman(2023)]{tuckerman2023statistical}
Mark~E Tuckerman.
\newblock \emph{Statistical mechanics: theory and molecular simulation}.
\newblock Oxford university press, 2023.

\bibitem[Unke and Meuwly(2018)]{unke2018reactive}
Oliver~T Unke and Markus Meuwly.
\newblock A reactive, scalable, and transferable model for molecular energies from a neural network approach based on local information.
\newblock \emph{The Journal of chemical physics}, 148\penalty0 (24):\penalty0 241708, 2018.

\bibitem[Unke et~al.(2021{\natexlab{a}})Unke, Chmiela, Gastegger, Sch{\"u}tt, Sauceda, and M{\"u}ller]{unke2021spookynet}
Oliver~T Unke, Stefan Chmiela, Michael Gastegger, Kristof~T Sch{\"u}tt, Huziel~E Sauceda, and Klaus-Robert M{\"u}ller.
\newblock Spookynet: Learning force fields with electronic degrees of freedom and nonlocal effects.
\newblock \emph{Nature communications}, 12\penalty0 (1):\penalty0 1--14, 2021{\natexlab{a}}.

\bibitem[Unke et~al.(2021{\natexlab{b}})Unke, Chmiela, Sauceda, Gastegger, Poltavsky, Sch\:utt, Tkatchenko, and M\"uller]{unke2021machine}
Oliver~T Unke, Stefan Chmiela, Huziel~E Sauceda, Michael Gastegger, Igor Poltavsky, Kristof~T Sch\:utt, Alexandre Tkatchenko, and Klaus-Robert M\"uller.
\newblock Machine learning force fields.
\newblock \emph{Chemical Reviews}, 121\penalty0 (16):\penalty0 10142--10186, 2021{\natexlab{b}}.

\bibitem[Van De~Walle and Ceder(2002)]{vandewalle2002effect}
Axel Van De~Walle and Gerbrand Ceder.
\newblock The effect of lattice vibrations on substitutional alloy thermodynamics.
\newblock \emph{Reviews of Modern Physics}, 74\penalty0 (1):\penalty0 11, 2002.

\bibitem[Wander et~al.(2024)Wander, Shuaibi, Kitchin, Ulissi, and Zitnick]{wander2024cattsunami}
Brook Wander, Muhammed Shuaibi, John~R Kitchin, Zachary~W Ulissi, and C~Lawrence Zitnick.
\newblock Cattsunami: Accelerating transition state energy calculations with pre-trained graph neural networks.
\newblock \emph{arXiv preprint arXiv:2405.02078}, 2024.

\bibitem[Wang et~al.(2020)Wang, Axelrod, and G{\'o}mez-Bombarelli]{wang2020differentiable}
Wujie Wang, Simon Axelrod, and Rafael G{\'o}mez-Bombarelli.
\newblock Differentiable molecular simulations for control and learning.
\newblock \emph{arXiv preprint arXiv:2003.00868}, 2020.

\bibitem[Weiler et~al.(2018)Weiler, Geiger, Welling, Boomsma, and Cohen]{weiler20183d}
Maurice Weiler, Mario Geiger, Max Welling, Wouter Boomsma, and Taco~S Cohen.
\newblock 3d steerable cnns: Learning rotationally equivariant features in volumetric data.
\newblock \emph{Advances in Neural Information Processing Systems}, 31, 2018.

\bibitem[Williams et~al.(2025)Williams, Kabalan, Stojanovic, Z{\'o}lyomi, and Pyzer-Knapp]{williams2025hessian}
Nicholas~J Williams, Lara Kabalan, Ljiljana Stojanovic, Viktor Z{\'o}lyomi, and Edward~O Pyzer-Knapp.
\newblock Hessian qm9: A quantum chemistry database of molecular hessians in implicit solvents.
\newblock \emph{Scientific Data}, 12\penalty0 (1):\penalty0 9, 2025.

\bibitem[Wines and Choudhary(2024)]{wines2024chips}
Daniel Wines and Kamal Choudhary.
\newblock Chips-ff: Evaluating universal machine learning force fields for material properties.
\newblock \emph{arXiv preprint arXiv:2412.10516}, 2024.

\bibitem[Yang et~al.(2024)Yang, Hu, Zhou, Liu, Shi, Li, Li, Chen, Chen, Zeni, et~al.]{yang2024mattersim}
Han Yang, Chenxi Hu, Yichi Zhou, Xixian Liu, Yu~Shi, Jielan Li, Guanzhi Li, Zekun Chen, Shuizhou Chen, Claudio Zeni, et~al.
\newblock Mattersim: A deep learning atomistic model across elements, temperatures and pressures.
\newblock \emph{arXiv preprint arXiv:2405.04967}, 2024.

\bibitem[Yin et~al.(2025)Yin, Wang, Du, Wang, Ying, Jia, Zhang, Du, Gomes, Duan, et~al.]{yin2025alphanet}
Bangchen Yin, Jiaao Wang, Weitao Du, Pengbo Wang, Penghua Ying, Haojun Jia, Zisheng Zhang, Yuanqi Du, Carla~P Gomes, Chenru Duan, et~al.
\newblock Alphanet: Scaling up local frame-based atomistic foundation model.
\newblock \emph{arXiv preprint arXiv:2501.07155}, 2025.

\bibitem[Zhang et~al.(2018)Zhang, Han, Wang, Saidi, Car, et~al.]{zhang2018end}
Linfeng Zhang, Jiequn Han, Han Wang, Wissam Saidi, Roberto Car, et~al.
\newblock End-to-end symmetry preserving inter-atomic potential energy model for finite and extended systems.
\newblock \emph{Advances in Neural Information Processing Systems}, 31, 2018.

\bibitem[Zitnick et~al.(2022)Zitnick, Das, Kolluru, Lan, Shuaibi, Sriram, Ulissi, and Wood]{zitnick2022spherical}
Larry Zitnick, Abhishek Das, Adeesh Kolluru, Janice Lan, Muhammed Shuaibi, Anuroop Sriram, Zachary Ulissi, and Brandon Wood.
\newblock Spherical channels for modeling atomic interactions.
\newblock \emph{Advances in Neural Information Processing Systems}, 35:\penalty0 8054--8067, 2022.

\bibitem[Zubatyuk et~al.(2019)Zubatyuk, Smith, Leszczynski, and Isayev]{zubatyuk2019accurate}
Roman Zubatyuk, Justin~S Smith, Jerzy Leszczynski, and Olexandr Isayev.
\newblock Accurate and transferable multitask prediction of chemical properties with an atoms-in-molecules neural network.
\newblock \emph{Science advances}, 5\penalty0 (8):\penalty0 eaav6490, 2019.

\end{thebibliography}
\bibliographystyle{assets/plainnat}

\newpage
\clearpage
\appendix
\renewcommand\thefigure{\thesection.\arabic{figure}}

\section{Experimental details}
\label{appendix:experimental}

\subsection{MD simulation protocol}
\label{appendix:conservation_protocol}

Simulating molecular systems not seen during training is a key capability of MLIPs. Therefore, we construct challenging MD simulation tasks featuring out-of-distribution data to test a model's conservation capability. We conduct experiments on two chemical domains: inorganic crystals and organic molecules.

\textbf{Inorganic materials.} For training models on inorganic materials we utilize the MPTrj~\citep{deng2023chgnet} dataset, and we establish a suite of simulation tasks based on the TM23 dataset~\citep{owen2024complexity}. TM23  contains MD samples of 27 single vacancy defect transition metal systems at cold, warm, and melt temperatures using an NVT ensemble, in total 81 combinations of different metals and temperature. These defected systems are out-of-distribution for a model trained on the MPTrj datasets, which only contains relaxation trajectories of non-defected systems. Additionally, some of the metals in the TM23 dataset is very rare in the MPTrj dataset. We initialize the simulation by sampling a frame from the TM23 dataset, run a relaxation using the LBFGS algorithm, randomly initialize the atom velocities at cold/warm/melt temperatures using a Maxwell-Boltzmann distribution, then run MD simulations under the NVE ensemble for 100 ps using a time step of 5 fs (same as the time step used in the TM23 ab initio MD protocol). 

\textbf{Organic molecules.} For training models on organic molecules we use the SPICE-MACE-OFF dataset~\citep{kovacs2023mace}, which is mainly based on the SPICE-1.0 dataset~\citep{eastman2023spice}, and we establish a suite of simulation tasks from the MD22 dataset~\citep{chmiela2023accurate} that contains seven large molecules. Molecules in MD22 are out-of-distribution for a model trained on the SPICE-MACE-OFF dataset as they are considerably larger than all molecules in the SPICE-MACE-OFF training dataset. We initialize the simulation by sampling a frame from the MD22 dataset, run a relaxation using the LBFGS algorithm, randomly initialize the atom velocities at a temperatures of 400/500 K using a Maxwell-Boltzmann distribution (400 K for Buckyball catcher and Double-walled nanotube and 500 K for other molecules, which are the same as the MD22 protocol), then run MD simulations under the NVE ensemble for 100 ps using a time step of 1 fs (same as the time step used in the MD22 ab initio MD protocol). 

All ML-based MD simulations use a Velocity-Verlet integrator and are conducted with \textsc{ASE}~\citep{ase}. We measure the energy conservation error (extent of energy drift) across the 100-ps simulations. All \ourmodel\ models are 2-layer with $L_{\mathrm{max}}=2$ and $M_{\mathrm{max}}=2$ (3.2M trainable parameters). Detailed model hyperparameters are included in \cref{appendix:hyperparams}.

\subsection{Phonon calculation protocols}

Harmonic and anharmonic phonon calculations and solutions to the Wigner transport equation \citep{simoncelli2022wigner} used for the thermal conductivity benchmark ($\kappa_{\text{SRME}}$) values given in \cref{tab:mbd} were carried out using the supercell method with finite differences implemented in \textsc{Phono3py} \citep{togo2015phono3py, togo2023phonopy}. The calculations followed the protocol described in the Matbench-Discovery benchmark~\citep{riebesell2023matbench, pota2024thermal}. For \ourmodel-30M-MP, an even lower $\kappa_{\mathrm{SRME}}$ of $0.298$ is obtained when we adjust the evaluation parameter atom displacement from $0.03$ \AA\ to $0.05$ \AA.

Harmonic phonon calculations for the MDR benchmark results listed in \cref{tab:mdr-phonons} were carried out following the calculation protocol used by \citep{loew2024universal} which employs phonon calculations using the supercell method with finite differences with a displacement of 0.01 \AA. Calculations were done using the \textsc{Phonopy} software~\cite{togo2023phonopy}.

\subsection{Test-set error for MPTrj-trained models} 

\begin{figure*}[t]
\includegraphics[width=\linewidth]{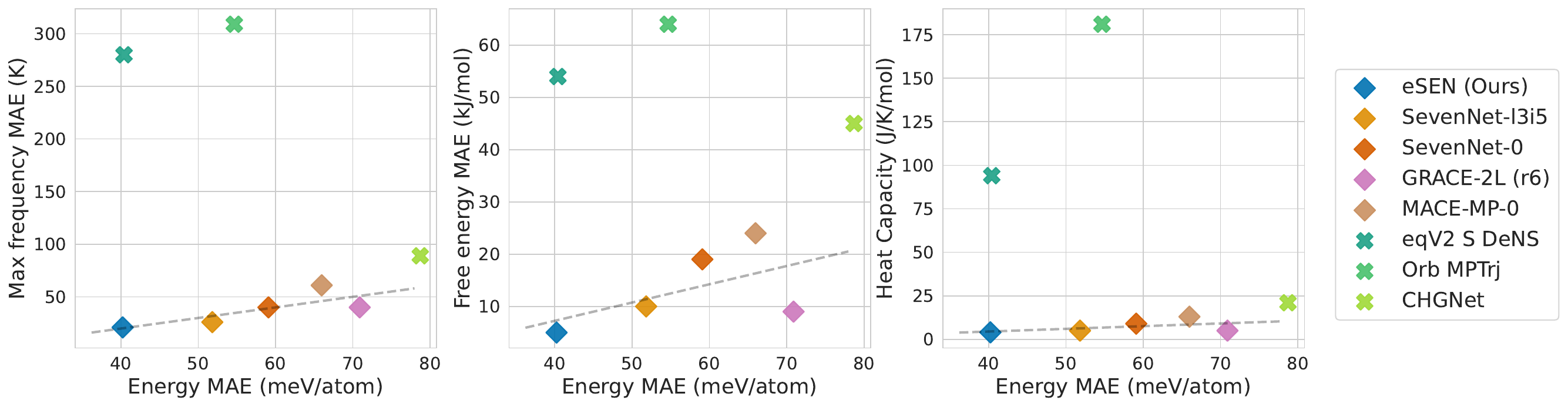}
\caption{The correlation between test-set energy error and maximum frequency, free energy and heat capacity across different model architectures.}
\label{fig:phonon_extra_inter}
\end{figure*}

\begin{figure*}[t]
\includegraphics[width=\linewidth]{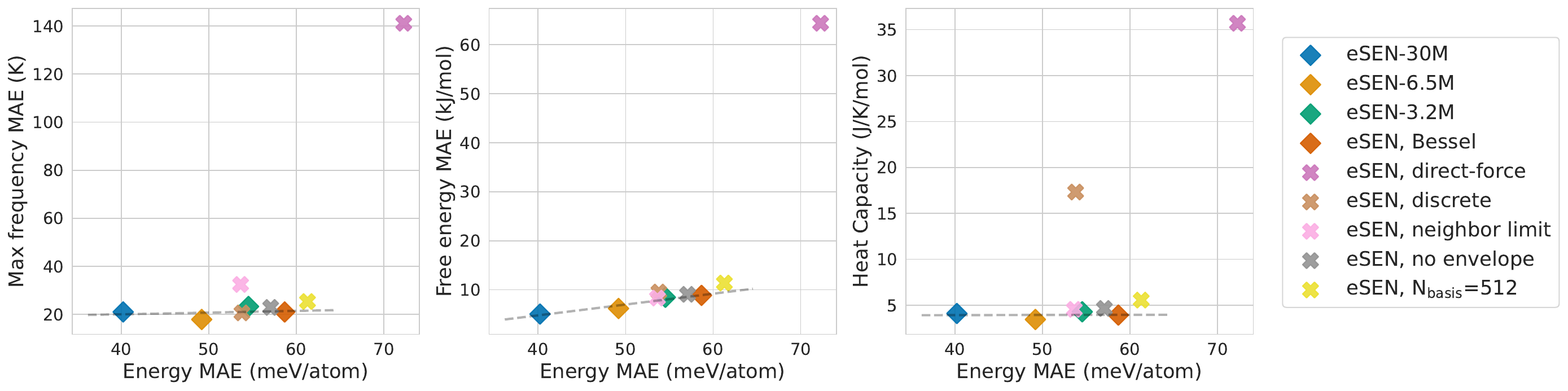}
\caption{The correlation between test-set energy error and maximum frequency, free energy and heat capacity for different variants of \ourmodel.}
\label{fig:phonon_extra_ours}
\end{figure*}

Since MPTrj lacks an official test split and various models are typically trained on distinct subsets of the data, we randomly selected 5000 samples from the subsampled Alexandria (sAlex) dataset~\citep{schmidt2024improving, barroso2024open} for a fair comparison. This subset was used to calculate the test-set energy mean absolute errors (MAEs) presented in Figures \ref{fig:overview}, \ref{fig:test_error_dev}, \ref{fig:phonon_extra_inter}, and \ref{fig:phonon_extra_ours}.

\section{Phonon calculations}\label{appendix:phonons}

\subsection{Correlation of test-set energy errors and vibrational property errors}

\cref{fig:phonon_extra_inter} presents the correlation between test-set energy MAE and the other three phonon calculation tasks, evaluated across various model architectures. The corresponding correlations for different variants of \ourmodel~are displayed in \cref{fig:phonon_extra_ours}. In both figures, improved correlation can be observed among energy-conserving models.

\cref{fig:phonon_extra_ours} illustrates that failing the conservation test can result in varying degrees of impact on different properties, depending on the specific design choices made. Although the neighbor limit, envelope function, and number of basis functions substantially affect $\kappa_{\mathrm{SRME}}$ (see \cref{fig:test_error_dev}), their influence on the properties evaluated in the MDR Phonon benchmark is relatively minor. Representation discretization impacts vibrational entropy and heat capacity but not other properties. While a model might still be able to get good performance in some physical property task when the energy conservation test is failed, when the conservation test is passed, the model performs very well robustly across all metrics. 

\subsection{Displacement values and their relation to phonon band structure predictions}

\begin{figure*}[t]
\includegraphics[width=\textwidth]{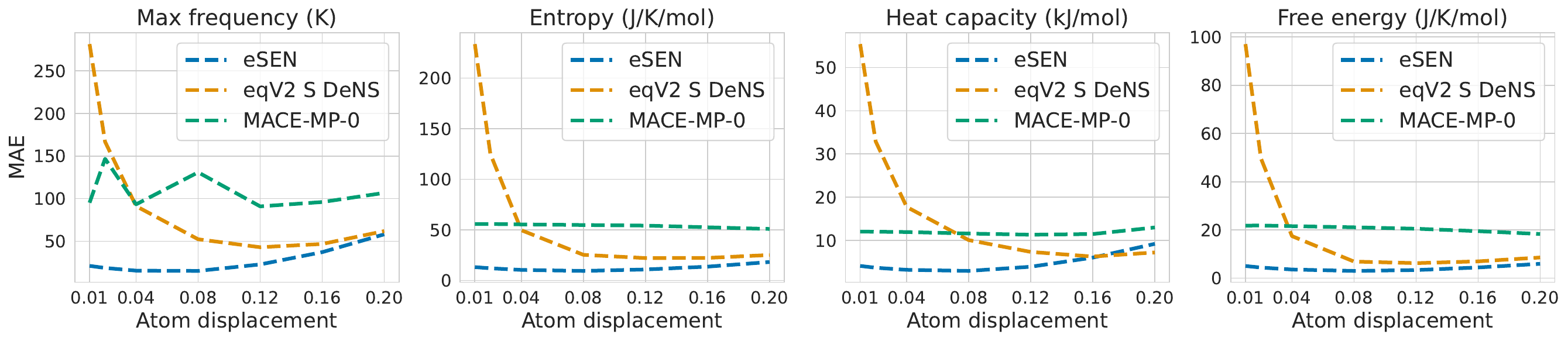}
\caption{
Errors in a randomly sampled subset (1000 samples) of the MDR Phonon benchmark when the atom displacement is adjusted.
}
\label{fig:phonon_atomdisp}
\end{figure*}

\cref{fig:phonon_atomdisp} presents the MAEs on phonon calculations for \ourmodel, MACE, and eqV2-S-DeNS as a function of increasing displacement values. As anticipated, both \ourmodel\ and MACE exhibit constant or slightly increasing MAE with respect to displacement. In contrast, eqV2-S-DeNS displays a notable decrease in MAE with increasing displacement. Notably, when using a displacement of 0.2 \AA, the resulting phonon benchmark MAE values for eqV2-S-DeNS become comparable to those of conservative force models (\cref{tab:mdr-phonons}). 

\begin{figure*}
\includegraphics[width=\textwidth]{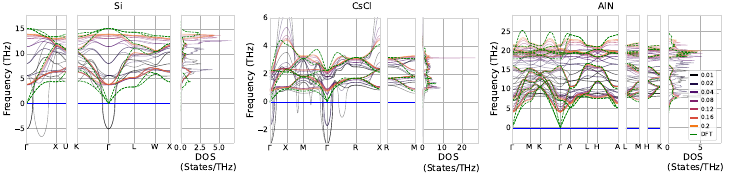}
\caption{Predicted phonon band structure and density of states (DOS) of Si (diamond structure), CsCl (CsCl structure), AlN (wurtzite structure) using eqV2-S-DeNS (direct-force prediction) at different displacement values. DFT baseline is taken from the PBE MDR dataset \cite{loew2024universal} calculated using a displacement of 0.01 \AA}
\label{fig:phonon_bands_eqV2}
\end{figure*}

While the prediction accuracy of thermodynamic properties such as maximum frequency, entropy, free energy, and heat capacity improves with increasing displacement for direct-force models, this improvement is deceptive and does not translate to accurate predictions of the underlying phonon band structure and density of states (DOS). As illustrated in \cref{fig:phonon_bands_eqV2}, the predicted phonon bands and DOS for three selected materials exhibit significant errors, particularly in capturing the correct dispersion relations. Moreover, imaginary frequencies are commonly predicted at small displacement values, suggesting a rough energy landscape (i.e. the learned PES is not truly convex when it's very close to the minima). The eqV2-S-DeNS model also fails to accurately capture acoustic modes---those that go to zero linearly at the $\gamma$ point---which is due to a non-zero net force on the structure. In contrast, non-zero net force at energy local minima does not occur for conservative models by definition (\cref{fig:phonon_bands}).

\begin{figure*}
\includegraphics[width=\textwidth]{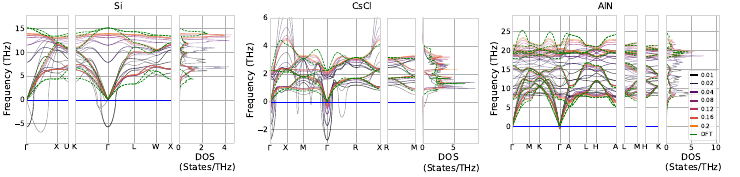}
\caption{Predicted phonon band structure and density of states (DOS) of Si (diamond structure), CsCl (CsCl structure), AlN (wurtzite structure) using eqV2-S-DeNS (direct-force prediction with a zero net force constraint) at different displacement values. DFT baseline is taken from the PBE MDR dataset \cite{loew2024universal} calculated using a displacement of 0.01 \AA}
\label{fig:phonon_bands_eqV2_zero_force}
\end{figure*}

By enforcing a net zero force prediction, as proposed by \citealt{neumann2024orb}, direct-force models can be modified to accurately capture acoustic phonon modes. As demonstrated in \cref{fig:phonon_bands_eqV2_zero_force}, incorporating this constraint allows the model to predict acoustic modes correctly. However, despite this improvement, the model continues to struggle with accurately reproducing the phonon band structure, and the issue of predicting imaginary frequencies at small displacement values persists.

The apparent paradox of direct-force models like eqV2-S-DeNS failing to accurately capture phonon band structures while still achieving competitive accuracy for thermodynamic properties such as entropy, free energy, and heat capacity can be resolved by examining the underlying calculation methodology. These properties are computed using weighted integrals of the DOS. Additionally the metrics in the MDR are performed at room temperature (300 K). The predicted DOS in \cref{fig:phonon_bands_eqV2} and \cref{fig:phonon_bands_eqV2_zero_force} at larger displacement values adequately captures the overall features of the DFT DOS, but does not reproduce finer details such as high density areas and fluctuations. This level of agreement is sufficient for accurate predictions because the Boltzmann-weighted integrals used in calculating thermal properties help to mitigate the impact of point-wise errors, making it less crucial to precisely capture fine details in the band structure and DOS \cite{ackland1997practical, vandewalle2002effect}. Moreover, since properties are estimated at 300 K, models can achieve accurate predictions by prioritizing prediction accuracy of lower-frequency modes, which are more relevant for thermal property calculations, rather than attempting to capture higher-frequency modes. Although eqV2-S-DeNS without a net-zero force contraint may not accurately capture acoustic phonon branches, its ability to predict vibrational thermodynamics at room temperature remains unaffected due to the relatively small number of acoustic phonon states at low frequencies.

\begin{figure*}
\includegraphics[width=\textwidth]{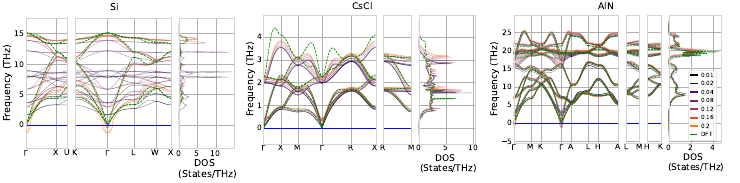}
\caption{Predicted phonon band structure and density of states (DOS) of Si (diamond structure), CsCl (CsCl structure), AlN (wurtzite structure) using \ourmodel~with direct-force prediction at different displacement values. DFT baseline is taken from the PBE MDR dataset \cite{loew2024universal} calculated using a displacement of 0.01 \AA}
\label{fig:phonon_bands_direct}
\end{figure*}

\begin{figure*}
\includegraphics[width=\textwidth]{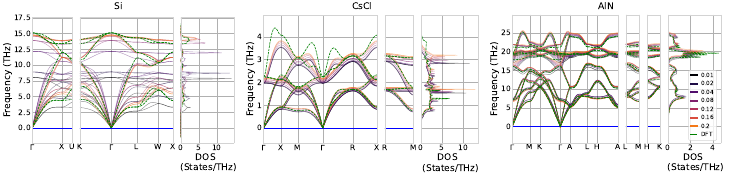}
\caption{Predicted phonon band structure and density of states (DOS) of Si (diamond structure), CsCl (CsCl structure), AlN (wurtzite structure) using \ourmodel\ with direct-force prediction and a zero-net force constraint at different displacement values. DFT baseline is taken from the PBE MDR dataset \cite{loew2024universal} calculated using a displacement of 0.01 \AA}
\label{fig:phonon_bands_direct_zero_force}
\end{figure*}

As a comparison with eqV2-S-DeNS, Figures \cref{fig:phonon_bands_direct} and \cref{fig:phonon_bands_direct_zero_force} shows the predicted phonon dispersion and DOS for the same three materials using \ourmodel\ with direct-force prediction. Although the results still exhibit some of the characteristic artifacts of direct-force models, such as convergence at larger displacements and the absence of acoustic modes, these issues are less pronounced compared to eqV2-S-DeNS. Moreover, the predicted phonon bands and DOS are significantly improved, providing a more accurate representation of the DFT reference values. The better approximation of phonon bands and a lower tendency to predict imaginary frequencies highlight the importance of a smoothly-varying model, even without being conservative.

Extending the existing metrics proposed by \citealt{loew2024universal} with additional evaluations would provide a more comprehensive assessment of MLIP performance. Specifically, new metrics could be developed to assess phonon dispersion across all modes and frequencies at commensurate points; and computing vibrational thermodynamic properties at a range of temperatures.

\begin{figure}[h]
\centering
\includegraphics[width=0.35\textwidth]{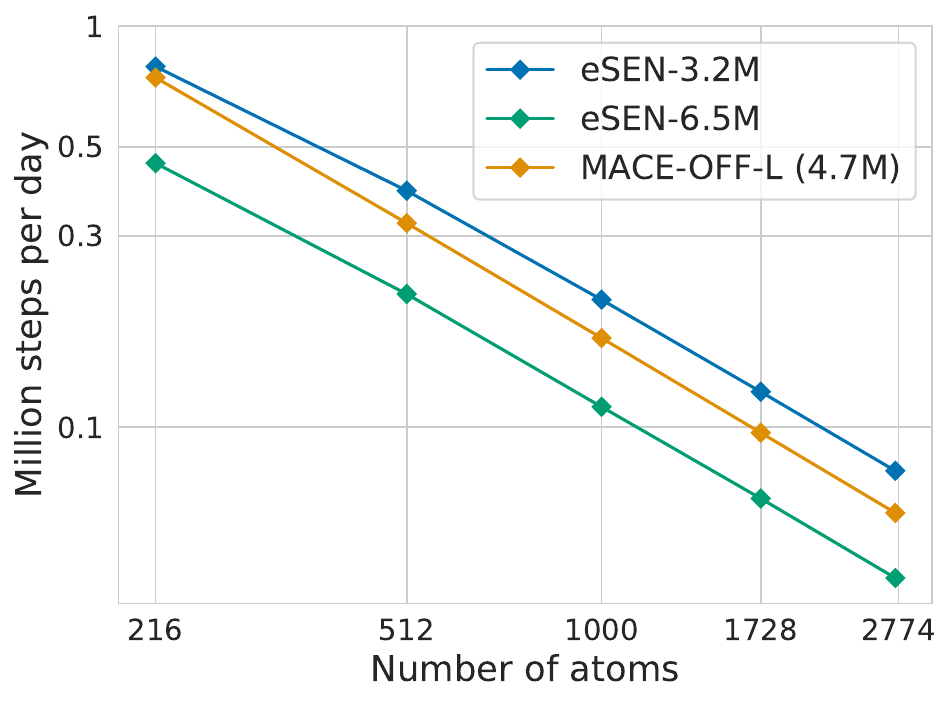}
\caption{Inference efficiency of MACE-OFF-L and \ourmodel s\ of a similar scale.}
\label{fig:inference}
\end{figure}

\section{Inference efficiency}
\label{appendix:inference}

We benchmarked the inference speed of our models against the similar sized MACE-OFF-L~\citep{kovacs2023mace} (4.7M) on a single 80GB Nvidia A-100 GPU. For MACE-OFF-L we used the exact benchmark code found in \url{https://github.com/ACEsuit/mace/blob/main/tests/test_benchmark.py}
 with mace-torch v0.3.6 (PyPi). To create a fair comparison, we replicated the identical benchmark environment as MACE benchmarks using the same diamond system with variable number of supercells (carbon atoms) as input. All models are benchmarked using the standard Python (3.12) runtime with Pytorch~v2.4.0~\citep{paszke2019pytorch} and CUDA~12.1~\citep{nickolls2008scalable}. No compile/torchscript was used for standardization of runtime. Across all system sizes, \ourmodel-3.2M has a comparable inference efficiency to MACE-OFF-L. For 216 atoms (\cref{fig:inference}), our models (3.2M, 6.4M) can run approximately (0.4, 0.8) million steps per day comparable to MACE-OFF-L (0.7 million steps per day).

\section{Hyper-parameters}
\label{appendix:hyperparams}

Hyper-parameters used for model training are shown in \cref{table:hps}. We train all models using a per-atom energy MAE loss, a force $l_2$ loss, and a stress MAE loss. For direct-force models or direct-force pre-training, we use the same decomposed loss as described in ~\citealt{barroso2024open}. The \ourmodel-30M model trained on MPTrj uses Denoising Non-equilibrium Structures (DeNS)~\citep{liao2024generalizing} with a noising probablity of $0.5$, a standard deviation of $0.1$ \AA\ for the added Gaussian noise, and $10$ for the DeNS loss coefficient during direct-force pre-training. DeNS is not used during OMat24 training or conservative fine-tuning. In our ablation study for maximum neighbor limit we used $30$ as the limit.

\begin{table*}[h]
\centering
\caption{Hyper-parameters for \ourmodel\ variants reported in this paper. \textsuperscript{*}\ourmodel-30M on MPTrj was trained for 60 epochs using direct-force pre-training and 40 epochs of conserved fine-tuning. \ourmodel-30M-OMat was trained for 2 epochs using  direct-force pre-training and 2 epochs of conserved fine-tuning. \textsuperscript{\textdagger}The \ourmodel-30M-OAM model starts from the  \ourmodel-30M-OMat model, and was finetuned for 1 epoch on a dataset constructed by combining the sAlex training dataset and 8 copies of the MPTrj training dataset.}
\label{table:hps}
\begin{adjustbox}{width=0.9\linewidth}
\begin{tabular}{l|ccccccc}
\toprule
Hyper-parameters & SPICE-3.2M & SPICE-6.5M & MPTrj-3.2M & MPTrj-6.5M & MPTrj-30M & OMat-30M & OAM Fine-tuning \\
\midrule
Number of \ourmodel~layer blocks & $2$ & $4$ & $2$ & $4$ & $10$ & $10$ & $10$ \\
Maximum degree $L_{\mathrm{max}}$ & $2$ & $2$ & $2$ & $2$ & $3$ & $3$ & $3$\\
Maximum order $M_{\mathrm{max}}$ & $2$ & $2$ & $2$ & $2$ & $2$ & $2$ & $2$\\
Number of channels $N_{\mathrm{channel}}$ & $128$ & $128$ & $128$ & $128$ & $128$ & $128$ & $128$ \\
Radial basis function & Bessel & Bessel & Gaussian & Gaussian & Gaussian & Gaussian & Gaussian \\
Number of radial basis functions & $10$ & $10$ & $10$ & $10$  & $10$ & $64$ & $64$ \\
Cutoff radius (\AA) & $5$ & $5$ & $6$ & $6$ & $6$ & $6$ & $6$ \\
Batch size & 128 & 128 & 512 & 512 & 512 & 512 & 256 \\
Optimizer & AdamW & AdamW & AdamW & AdamW & AdamW & AdamW & AdamW \\
Learning rate scheduling & Cosine & Cosine & Cosine & Cosine & Cosine & Cosine & Cosine \\
Warmup epochs & $0.1$ & $0.01$ & $0.1$ & $0.1$ & $0.1$ & $0.1$ & $0.1$ \\
Warmup factor & $0.2$ & $0.2$ & $0.2$ & $0.2$ & $0.2$ & $0.2$ & $0.2$  \\
Maximum learning rate & $4 \times 10 ^{-4}$ & $4 \times 10^{-4}$ & $4 \times 10 ^{-4}$ & $4 \times 10^{-4}$ & $4 \times 10 ^{-4}$ & $4 \times 10 ^{-4}$ & $2 \times 10 ^{-4}$ \\
Number of epochs & $100$ & $100$ & $100$ & $100$ & $60 + 40$\textsuperscript{*} & $2 + 2$\textsuperscript{*} & $1$\textsuperscript{\textdagger} \\
Gradient clipping norm & $100$ & $100$ & $100$ & $100$ & $100$ & $100$ & $100$ \\
Model EMA decay & $0.999$ & $0.999$ & $0.999$ & $0.999$& $0.999$ & $0.999$ & $0.999$ \\
Weight decay & $1 \times 10 ^{-3}$ & $1 \times 10 ^{-3}$ & $1 \times 10 ^{-3}$ & $1 \times 10 ^{-3}$ & $1 \times 10 ^{-3}$ & $1 \times 10 ^{-3}$ & $1 \times 10 ^{-3}$ \\
Energy loss coefficient & $10$ & $10$ & $1$ & $1$ & $20$ & $20$ & $20$ \\
Force loss coefficient & $20$ & $20$ & $10$ & $10$ & $20$ & $20$ & $20$\\
Stress loss coefficient & - & - & $100$ & $100$ & $5$ & $5$ & $5$ \\
\bottomrule
\end{tabular}
\end{adjustbox}
\end{table*}

\end{document}